\documentclass[proof]{pasj01}
\usepackage{subfig} 
\usepackage{graphicx} 

\draft 
\Received{$\langle$reception date$\rangle$}
\Accepted{$\langle$acception date$\rangle$}
\Published{$\langle$publication date$\rangle$}

\begin{document}
\title{A comparative study of amplitude calibrations for East-Asia VLBI Network: a-priori and template spectrum methods}

\author{Ilje \textsc{Cho}\altaffilmark{1,2}}
\author{Taehyun \textsc{Jung}\altaffilmark{1,2}$^{\rm *}$}
\author{Guang-Yao \textsc{Zhao}\altaffilmark{1}}
\author{Kazunori \textsc{Akiyama}\altaffilmark{3,4,5}}
\author{Satoko \textsc{Sawada-Satoh}\altaffilmark{6,7}}
\author{Motoki \textsc{Kino}\altaffilmark{12,1,5}}
\author{Do-Young \textsc{Byun}\altaffilmark{1,2,8}}
\author{Bongwon \textsc{Sohn}\altaffilmark{1,2,8}}
\author{Katsunori M. \textsc{Shibata}\altaffilmark{5,9}}
\author{Tomoya \textsc{Hirota}\altaffilmark{5,9}}
\author{Kotaro \textsc{Niinuma}\altaffilmark{10}}
\author{Yoshinori \textsc{Yonekura}\altaffilmark{7}}
\author{Kenta \textsc{Fujisawa}\altaffilmark{10,11}}
\author{Tomoaki \textsc{Oyama}\altaffilmark{5}}

\altaffiltext{1}{Korea Astronomy and Space Science Institute, Daedeok$-$daero 776, Yuseong$-$gu, Daejeon 34055, Korea}
\altaffiltext{2}{University of Science and Technology, Gajeong$-$ro 217 , Yuseong$-$gu, Daejeon 34113, Korea}
\altaffiltext{3}{Massachusetts Institute of Technology, Haystack Observatory, Route 40, Westford, MA 01886, USA}
\altaffiltext{4}{Black Hole Initiative, Harvard University, 20 Garden Street, Cambridge, MA 02138, USA}
\altaffiltext{5}{National Astronomical Observatory of Japan, Osawa 2-21-1, Mitaka, Tokyo 181-8588, Japan}
\altaffiltext{6}{Graduate School of Science and Engineering, Kagoshima University, 1-21-35 Korimoto, Kagoshima 890-0065, Japan}
\altaffiltext{7}{Center for Astronomy, Ibaraki University, 2-1-1 Bunkyo, Mito, Ibaraki 310-8512, Japan}
\altaffiltext{8}{Yonsei University, Yonsei-ro 50, Seodaemun-gu, Seoul 03722, Korea}
\altaffiltext{9}{Department of Astronomical Sciences, SOKENDAI (The Graduate University for Advanced Studies), Osawa 2-21-1, Mitaka, Tokyo 181-8588, Japan}
\altaffiltext{10}{Graduate School of Sciences and Technology for Innovation, Yamaguchi University, 1677-1 Yoshida, Yamaguchi, Yamaguchi 753-8511, Japan}
\altaffiltext{11}{The Research Institute for Time Studies, Yamaguchi University, 1677-1 Yoshida, Yamaguchi, Yamaguchi 753-8511, Japan}
\altaffiltext{12}{Academic Support Center, Kogakuin University, 2665-1 Nakano, Hachioji, Tokyo 192-0015, Japan}

\email{thjung@kasi.re.kr}


\maketitle

\begin{abstract}
We present the results of comparative study of amplitude calibrations for East-Asia VLBI Network (EAVN) at 22 and 43 GHz using two different methods of an ``a-priori" and a ``template spectrum", particularly on lower declination sources. 
Using observational data sets of early EAVN observations, we investigated the elevation-dependence of the gain values at seven stations of the KaVA (KVN and VERA Array) and three additional telescopes in Japan (Takahagi 32m, Yamaguchi 32m and Nobeyama 45m). By comparing the independently obtained gain values based on these two methods, we found that the gain values from each method were consistent within 10\% at elevations higher than 10$^\circ$. We also found that the total flux densities of two images produced from the different amplitude calibrations were in agreement within 10\% at both 22 and 43 GHz. By using the template spectrum method, furthermore, the additional radio telescopes can participate in the KaVA (i.e. EAVN) so that it can give a notable sensitivity increase. 
Therefore, our results will constrain the detailed conditions to reliably measure the VLBI amplitude using EAVN and give a potential to extend possible telescopes comprising EAVN. 
\end{abstract}

\section{Introduction \label{sec:intro}}
Recently, and thanks to the remarkable technical advances in very long baseline interferometry (VLBI), high frequency (e.g. $>$ 20 GHz) VLBI networks have been more generalized, for example Global Millimeter VLBI Array (GMVA), Event Horizon Telescope (EHT), Korean VLBI Network (KVN) and VLBI Exploration of Radio Astrometry (VERA). (Sub) milli-arcsecond (mas) scale of angular resolution can be easily achieved with these arrays and reveal the detailed structures of radio sources, particularly in the inner regions of active galactic nuclei (AGN) by overcoming opacity barrier. 
Flux densities and sizes of the VLBI structure based on an accurate measurement of visibility amplitude bring critical constraints on the physical properties of the radio cores and jets in AGN responsible for their emission mechanism, kinematics, variability, and magnetic field properties (e.g.  
\cite{doeleman:2008}, \yearcite{doeleman:2012}; \cite{fish:2011, lu:2011, akiyama:2015, johnson:2015, hada:2016, leej:2016}). \\
The KVN and VERA Array (KaVA) is a dedicated VLBI network of which regular open-use observations started in 2014 at 22 and 43 GHz. It consists of seven radio telescopes (four telescopes are from VERA, and three telescopes are from KVN) with maximum baseline length of $\sim$2,300 km in Korea and Japan. Thus the achievable angular resolution is up to approximately 1.2 and 0.6 mas at 22 and 43 GHz, respectively. 
Furthermore, international efforts on extending KaVA to the East Asian regions have been started since 2009 (\cite{miyazaki:2009}), named as the East Asia VLBI Network (EAVN; e.g. \cite{hagiwara:2015, wajima:2016}) by involving other telescopes from the Japanese VLBI Network (JVN; \cite{doi:2006} and references therein), and the Chinese VLBI Network (CVN; \cite{zheng:2015} and references therein). In particular, participations of large telescopes (e.g. Nobeyama 45~m telescope in Japan and Tianma 65~m telescope in China) will dramatically improve the sensitivity of the entire array. Although the EAVN has been in an early stage of commissioning, some demonstrative observations have been made by combining KaVA with additional telescopes such as the Nobeyama 45~m telescope, Takahagi 32~m telescope and Yamaguchi 32~m telescope in Japan. \\
However, an accurate calibration of VLBI visibility amplitude at these frequencies still remains as a difficult challenge, since temporal changes of sky opacity and elevation dependent gain variations become dominant (\cite{moran:1995}). In particular, mm/sub-mm VLBI observations experience adverse conditions because the image sensitivity is often limited by severe atmospheric environments. The most widely-used VLBI amplitude calibration is ``a-priori" calibration to measure the system equivalent flux density (SEFD) based on the opacity-corrected system temperature (T$_{sys}^\ast$) measurements and the elevation dependent gain curve for each telescope, which is equivalent to the auto-correlation spectrum and therefore can be used to scale the visibility amplitude (e.g. \cite{marti-vidal:2012}; see Section \ref{sec:reduction} and KaVA status report\footnote{http://radio.kasi.re.kr/kava}). In some stations of EAVN, however, the a-priori calibration is not applicable due to the lack of SEFD information. Alternatively, amplitude calibration using a planet (e.g. \cite{krichbaum:1993a}, \yearcite{krichbaum:1993b}) or the relative gain acquisition using an ambient maser source (so called ``template spectrum" calibration) is useful when instrumental effects are relatively unstable, usually for high frequency VLBI observations, to determine the telescope gain precisely (e.g. \cite{doeleman:2001, shen:2005, niinuma:2014}). \\
Therefore, we analyzed the amplitude calibrations using the a-priori and template spectrum methods at both 22 and 43 GHz, particularly for observations of lower declination sources at lower elevation (i.e. $\lesssim$ 40$^\circ$) where telescope performance is degraded and atmospheric opacity effects become severe. To compare performance between the a-priori and template spectrum methods, careful selection of the maser source is important. A fair comparison can be made from a maser source located spatially close (a few degrees) to the target source, with the total-power spectrum used as a reference for direct SEFD measurement. Our galactic center, Sagittarius A* (Sgr A*), is one of the best sources for this study, because there are strong H$_{2}$O and SiO maser sources in its neighbors (e.g. \cite{reid:2004, oyama:2008}; see Table~\ref{tab:soulist}). In addition, Sgr A* is one of the main sources in the KaVA long term monitoring program, so its accurate flux density measurement has great importance and is also crucial for its size estimation (e.g. \cite{lu:2011}; \cite{akiyama:2013}, \yearcite{akiyama:2014}). \\
This paper focuses on comparing gain estimation from the different amplitude calibration methods adding other telescopes outside KaVA. Detailed study of Sgr A* will be discussed in forthcoming papers. Some preliminary results from these VLBI experiments have already been reported (e.g. \cite{akiyama:2014}, \cite{zhao:2017}). Section \ref{sec:obs} outlines the observational details, and Section \ref{sec:reduction} describes the two amplitude calibration methods and data reduction procedures. Section \ref{sec:results} and \ref{sec:discussions} present the analysis results and discussions, respectively. Section \ref{sec:conclusion} presents our conclusions. \\

\begin{table}
\tbl{Source list.}{%
\centering
\begin{tabular}{llllll}
\hline
Sources & Sgr A* & NRAO 530 & Sgr B2 & OH 0.55-0.06 & VX Sgr \\
\hline
R.A. & 17:45:40.03 & 17:33:02.70 & 17:47:20.18 & 17:47:08.96 & 18:08:04.04 \\
Dec. & -29:00:28.0 & -13:04:49.5 & -28:23:03.88 & -28:29:55.46 & -22:13:26.61 \\
Maser lines & - & - & H$_2$O & SiO (J=1-0, v=1, 2) & SiO (J=1-0, v=1, 2) \\
$\theta_{sep}$$^{\rm a}$ (deg) & - / 16.19 & 16.19 / - & 0.72 / 15.66 & 0.60 / 15.77  & 8.45 / 12.37  \\
\hline
\end{tabular}}
\label{tab:soulist}
\begin{tabnote}
$^{\rm a}$ The separation angle from Sgr A* / NRAO 530. \\
\end{tabnote}
\end{table}

\begin{table}
\tbl{Observations list.}{%
\centering
\begin{tabular}{llll}
\hline
Project code & r13083b (K13) & r13102a (Q13) & r14308a (Q14) \\
\hline
Obs. Date & Mar 28, 2013 & Apr 12, 2013 & Nov 04, 2014  \\
Frequency & 22 GHz & 43 GHz & 43 GHz \\
Telescopes$^{\rm a}$ & KaVA, TAK$^{\rm b}$, YAM$^{\rm c}$ & KaVA, NRO & KaVA  \\
Maser source & Sgr B2 & OH 0.55-0.06, VX Sgr & OH 0.55-0.06, VX Sgr  \\
Reference telescope & KUS & KUS & KYS  \\
Reference time-range & UT 21:32 - 21:33 & UT 19:37 - 19:38 & UT 07:09 - 07:10 \\
\hline
\end{tabular}}
\label{tab:obslist}
\begin{tabnote}
$^{\rm a}$ KaVA : KYS (KVN Yonsei, 21m), KUS (KVN Ulsan, 21m), KTN (KVN Tamna, 21m), MIZ (Mizusawa, 20m), IRK (Iriki, 20m), OGA (Ogasawara, 20m), and ISG (Ishigaki-jima, 20m). Additional telescopes are TAK (Takahagi, 32m), YAM (Yamaguchi, 32m) and NRO (Nobeyama radio observatory, 45m). \\
$^{\rm b}$ TAK aperture efficiency was assumed as 30\%. \\
$^{\rm c}$ YAM was excluded from further analysis by severe weather condition. \\
\end{tabnote}
\end{table}

\section{Observation \label{sec:obs}}
For this purpose, three KaVA observations for Sgr A* from 2013 to 2014 were selected: one 22 GHz observation in March 28, 2013 (K13), and two 43 GHz observations in April 12, 2013 (Q13) and November 4, 2014 (Q14). There was one month difference between K13 and Q13, and three JVN telescopes also participated: the Takahagi 32m (TAK) and Yamaguchi 32m (YAM) telescopes participated in K13 even if YAM failed to detect the source due to bad weather, and the Nobeyama radio observatory (NRO, 45m) participated in Q13. 
The maximum aperture efficiency of TAK was reasonably assumed as 30\% (\cite{yonekura:2016}). Q14 was observed with KaVA only under a good weather conditions, so the data provides a good opportunity to check consistency (see Table~\ref{tab:obslist}). \\
While pointing offset measurement and correction were not applied to VERA telescopes and TAK, KVN measured and corrected it every 1-2 hours using a nearby continuum source (e.g. NRAO 530). For both KVN and VERA, typical pointing accuracy was better than 10 arcsec. NRO also corrected pointing offset every 1-2 hours using nearby maser sources and the offset was a few to 12 arcsec under a good weather conditions with slow wind speed less than 2 m/s during the observing time. \\
Observations were performed for 6 hours including multiple scans for calibrators together with Sgr A*. NRAO 530 was observed as a fringe finder, as well as the gain and delay calibrator for each observation. Maser lines of H$_{2}$O from Sgr B2 at 22 GHz, and SiO from OH 0.55-0.06 and VX Sgr at 43 GHz were used for the template spectrum method. Rest frequencies of H$_{2}$O and SiO maser lines were 22.235080, 42.820587 (v=2, J=1-0), and 43.122080 (v=1, J=1-0) GHz, respectively. \\
The total observing bandwidth was 256 MHz (16 MHz$\times$16 IFs) at 1 Gbps sampling rate. All data were correlated at the Korea-Japan Correlation Center in Daejeon, Korea (KJCC; \cite{lee:2015a}) with accumulation period 1.6 seconds. \\

\section{Data reduction \label{sec:reduction}}
\subsection{Initial calibrations \label{sec:initial_cal}}
All data were processed using the NRAO Astronomical Imaging Processing System (AIPS; \cite{greisen:2003}). Cross-correlation spectra were normalized by total-power spectra using the AIPS task, ACCOR. Bandpass calibration (AIPS task: BPASS) was also applied using total-power spectra of NRAO 530. \\
For phase calibrations, the total electron content (TEC) was corrected using the global TEC map provided by the Jet Propulsion Laboratory (AIPS task: VLBATECR). The parallactic angle was corrected by the AIPS task, VLBAPANG. To remove instrumental phase offsets between intermediate frequency (IF) channels, manual phase calibration was performed using NRAO 530 over a short time range (30 sec) where the signal-to-noise ratio (SNR) was highest, which enabled to integrate all channels for fringe search so that it increases the sensitivity (AIPS task: FRING). Solution intervals were 60 and 30 sec within atmospheric coherence time at 22 and 43 GHz, respectively. Consequently, Sgr A* was detected with a few tens of SNR at all telescopes except for ISG, YAM in K13 and OGA in Q13, as well as NRAO 530. \\

\subsection{Amplitude calibrations \label{sec:ampcal}}
For VLBI observations, visibility amplitude, $|V_{ij}|$, is obtained from the correlation coefficient, $C_{ij}$, when the source contribution to total-power spectra is negligible compared to SEFD (Jy),
\begin{eqnarray}
\label{eq:vis}
C_{ij} = \frac{|V_{ij}|}{\sqrt{\rm SEFD_i} \sqrt{\rm SEFD_j}},
\end{eqnarray}
where the suffixes $i$ and $j$ represent the telescopes comprising the baseline. Amplitude calibration calculates square-root values of ${\rm SEFD_i}$, so called gain solutions ($G_i$). This study compared two different amplitude calibration methods as described in detail in the following subsections. 

\subsubsection{A-priori method \label{sec:a-priori}}
A-priori method (AIPS task: APCAL) calculates the time varying gain solutions, $G_{i}^{AP}$ (Jy$^{1/2}$), from the opacity-corrected system temperature ($T_{sys}^{*}$) and aperture efficiency depending on the elevation of observing sources, so called gain curve: 
\begin{eqnarray}
\label{eq:gainsoltru}
G_{i}^{AP}(t) = \sqrt {\frac{1}{\rm DPFU_i} \frac{T_{sys,i}^{*}(t)}{f_i(e(t))}}, 
\end{eqnarray}
where $e(t)$ is target elevation at time $t$ and $f_{i}(e(t))$ is a dimensionless gain curve of $i$-th telescope that represents a fractional variation in the aperture efficiency normalized by the zenith aperture efficiency. 
${\rm DPFU}$ (= $A_{e}/2k$) stands for degrees  per flux density unit (K/Jy) where $A_e$ is the effective aperture size to the zenith and $k$ is the Boltzmann constant, so that it gives the temperature measurement sensitivity of an effective aperture area. Based on the $A_e$ which has been measured every year (see KaVA status report$^{1}$), therefore, the ${\rm DPFU}$ is given as a constant and its elevation dependent changes are given by the gain curve ($f_{i}(e(t))$). Note that the $A_e$ variation by the antenna pointing offset has been ignored, and the gain uncertainties are shown in Appendix A. While KVN uses a gain curve as the second-order polynomial function normalized by the value at zenith, VERA uses a constant value of unity. \\
Accurate $T_{sys}^{*}$ measurement and gain curve information are important for VLBI amplitude calibration. The $T_{sys}^{*}$ contains systematic errors of 5-10\% in its measurements using chopper-wheel calibrations, and the gain curves have $\sim$5\% of uncertainties for higher elevations ($>$40$^\circ$) according to the annual KaVA status report (see Appendix A, for typical errors at each telescope in our observations).
However, for lower elevation sources, the measured $T_{sys}^{*}$ has more ground spill-over and the sky temperature even if the use of chopper-wheel deals with the opacity effects to the first-order approximation. In addition, the uncertainties in aperture efficiency become larger for lower antenna elevations due to the difficulties to trace gravitational deformation accurately. Therefore, these errors affect the gain solutions of the a-priori method when observing sources at lower elevations. 

\begin{figure*}
\begin{center}
\includegraphics[height=100mm,width=100mm]{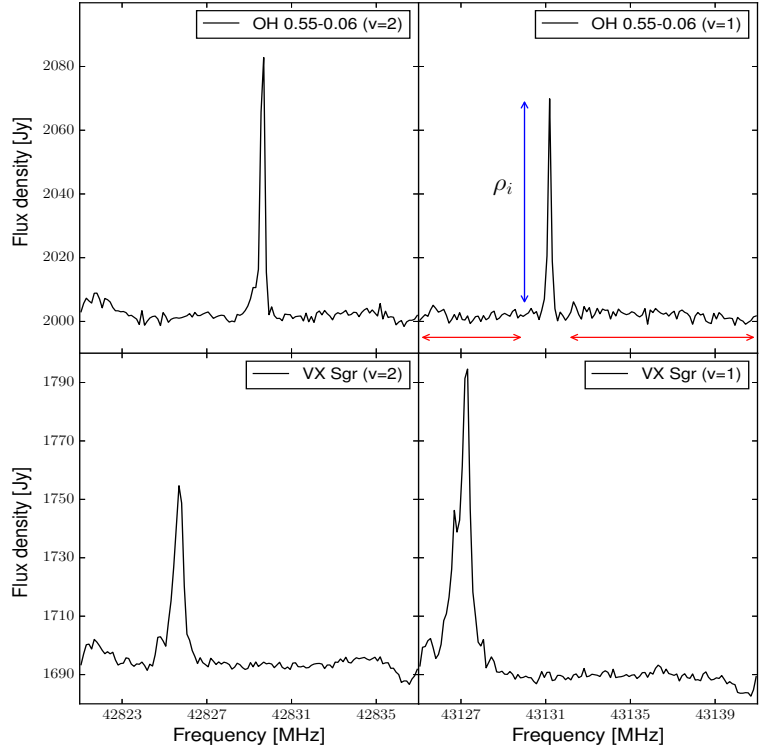}
\end{center}
\caption{The total-power spectra of SiO maser lines from each source which were observed in KUS and scan-averaged (for 2 minutes): (upper-left) v=2, J=1-0, (upper-right) v=1, J=1-0 from OH 0.55-0.06, (lower-left) v=2, J=1-0, (lower-right) v=1, J=1-0 from VX Sgr. Bandpass calibration through NRAO 530 was applied. Note that the flux densities were obtained through a-priori amplitude calibration. The $\rho$ measurement through a bright spectral line of maser source is shown with blue-arrow. The red-arrows are the baseline range of total-power spectrum of a maser. \label{fig:rho}}
\end{figure*}
%
\begin{figure*}
\centering
\includegraphics[height=70mm,width=0.355\textwidth]{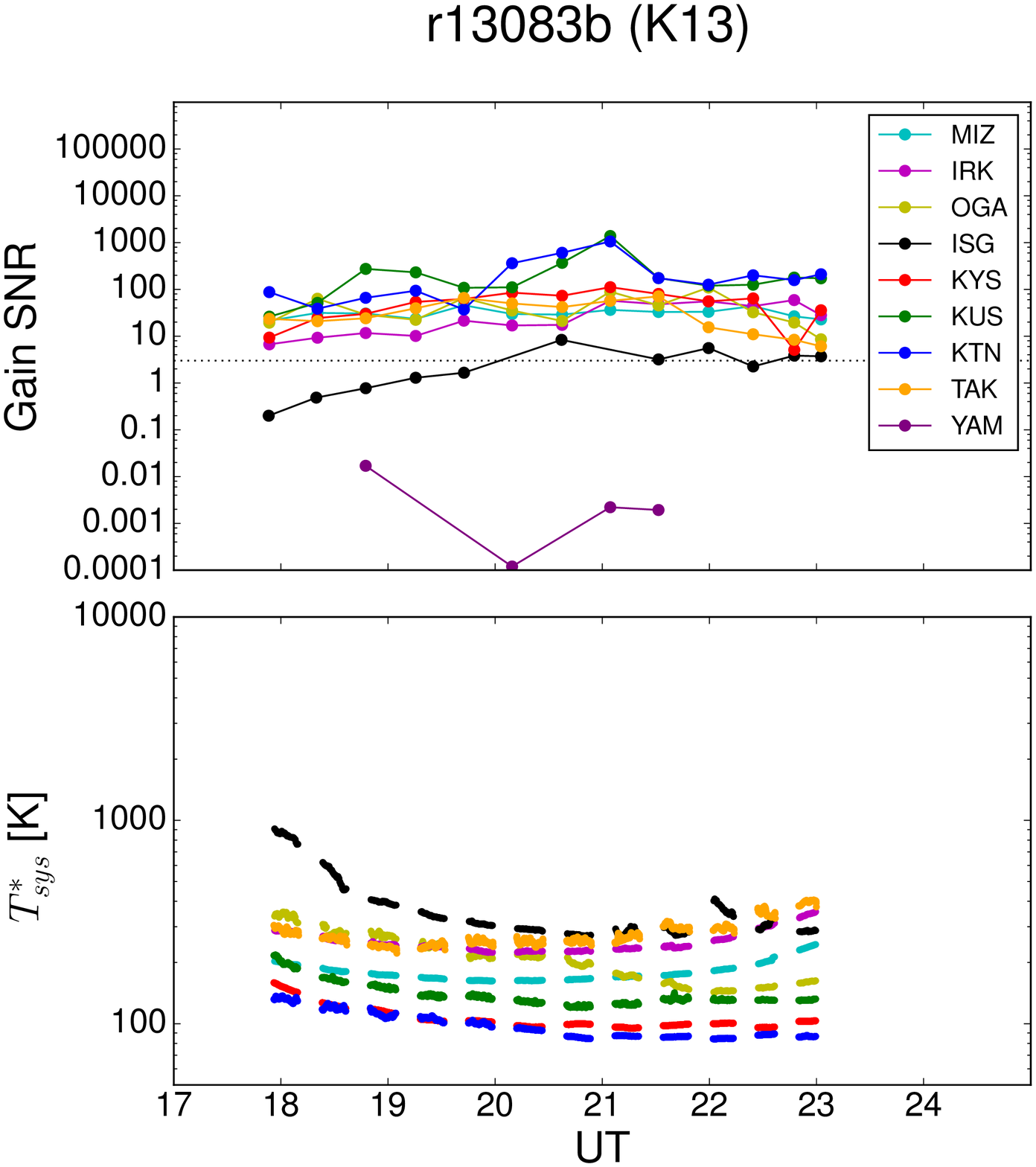}
\includegraphics[height=70mm,width=0.31\textwidth]{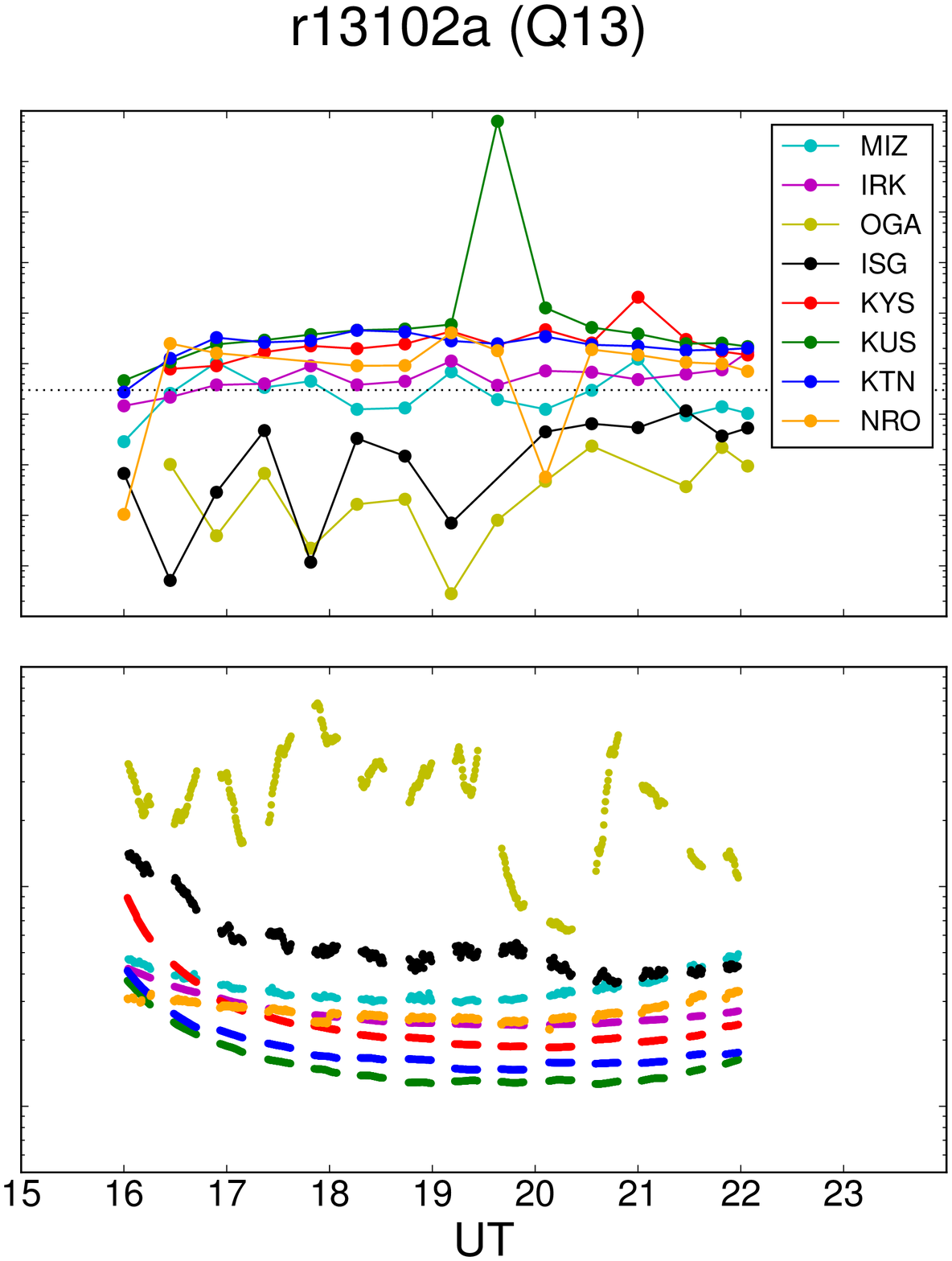}
\includegraphics[height=70mm,width=0.31\textwidth]{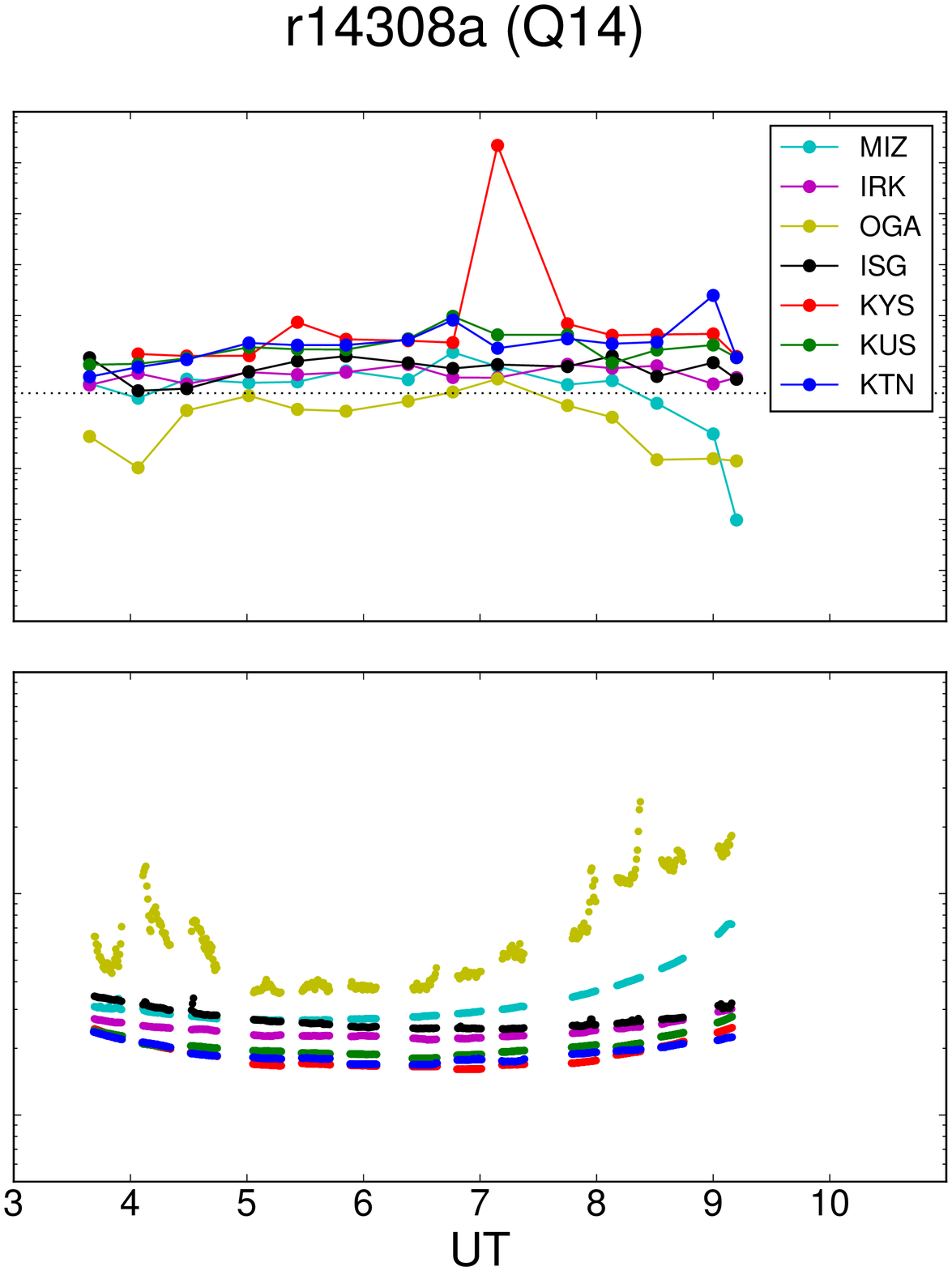}
\caption{(Upper) The signal-to-noise ratio (SNR) of gain solutions from template spectrum method which indicates the $\rho_i(t)$. (Lower) Opacity-corrected system temperature ($T_{sys}^{*}$) for each dataset: (left) r13083b, (middle) r13102a, and (right) r14308a. Each color represents each telescope as same for both SNR and $T_{sys}^{*}$. Note that the $T_{sys}^{*}$ of YAM was missed because no data was available (left panel, lower plot). The dotted line shows SNR = 3 which can be the lower-limit of clear detection of a maser line so that it can give reliable solutions.  \label{fig:acfitsnr}}
\end{figure*}

\subsubsection{Template spectrum method \label{sec:templatemethod}}
Template spectrum method (AIPS task: ACFIT) obtains the relative gain values using the maser source total-power spectra, as shown in Figure~\ref{fig:rho}. A high-quality spectrum is selected from a scan with a sensitive antenna at a reasonable elevation (i.e. `template') and is fitted to the total-power spectra from all telescopes using a linear least-squares algorithm. These can be simply translated to absolute gains by multiplying the measured $T_{sys}^{*}$ and aperture efficiency of the `template' so the gain solution, $G_{i}^{AC}$ (Jy$^{1/2}$) , can be expressed as,
\begin{eqnarray}
\label{eq:gainsol}
G_{i}^{AC}(t) = \sqrt{\frac{1}{\rm DPFU_0} \frac{T_{sys,0}^{*}(t_0)}{f_0(e(t_0))} \frac{\rho_0(t_0, \theta_{0}(t_0))}{\rho_i(t, \theta_{i}(t))}},
\end{eqnarray}
where $\theta(t)$ is the pointing offset at time $t$, and the suffix 0 means the reference telescope and time for a template usually when its gain is at the highest level. The $\rho$ is the normalized difference between the maser peak and its baseline spectrum (i.e. $\rho \equiv (P_{peak} - P_{base})/P_{base}$, where $P_{peak}$ is the total-power spectrum at the maser peak and $P_{base}$ is the mean total-power of its baseline; see Figure~\ref{fig:rho}), and the peak line flux density is assumed to be constant during the observation. Therefore, the $\rho$-ratio in Equation (\ref{eq:gainsol}) corresponds to, 
\begin{eqnarray}
\label{eq:rhoratio}
\frac{\rho_{0}(t_0)}{\rho_{i}(t)} = \frac{P_{peak,0}(t_0) - P_{base,0}(t_0)}{P_{peak,i}(t) - P_{base,i}(t)} \frac{P_{base,i}(t)}{P_{base,0}(t_0)}.
\end{eqnarray}
Thus it indicates the relative SEFD for each telescope to the template spectrum at the reference telescope and reference scan. In practice, the first and second terms on the right-hand side of Equation (\ref{eq:rhoratio}) correspond to the ratio of standard-deviation of total-power spectra to the template spectrum and their cross-correlation, respectively, as implemented in AIPS. Note that the $\rho$-ratio can correct the pointing offset at each time and telescope by assuming $\theta_{0}(t_0) \approx$ 0, so that $\rho_0(t_0, \theta_{0}(t_0)) / \rho_i(t, \theta_{i}(t)) \sim \rho_0(t_0) / \rho_i(t) \eta_{\theta,i}(t)$, where $\eta_{\theta,i}(t)$ is the gain efficiency when pointing offset is given. \\
Template spectrum method can provide a better estimate of antenna gain compared to a-priori method if a good template spectrum is used, particularly for observations of lower elevation sources. However, additional uncertainties can be introduced in the $\rho$ measurements, such as maser line detection, spatial distribution of a maser source and a template spectrum selection. Note that the bandpass calibration should be properly done before applying the template spectrum method to normalize the spectral baselines at each station, which may have different amplitude bandpass responses. To minimize the uncertainties, the following considerations are required:  1) the maser lines should be clearly detected over SNR $\sim$ 3 at least (see Figure~\ref{fig:acfitsnr}) where the SNR indicates $\rho_i(t)$, 2) maser sources should have a narrow and strong peak with spatially compact distribution compared to the telescope beam size, 3) a reference scan for a template spectrum should be selected with reasonable source elevation ($>$ 10$^\circ$) with relatively good aperture efficiency, and 4) a reference telescope for a template spectrum should have the clearest total-power spectrum of a maser line without ambiguous peaks or fluctuating baseline. We will discuss these effects in Section \ref{sec:discussions}. \\
For Sgr A* observations, one H$_{2}$O maser line from Sgr B2 was available for the template spectrum method at 22 GHz, and four SiO maser lines from OH 0.55-0.06 (v=2, 1, J=1-0) and VX Sgr (V=2, 1, J=1-0) were tested for amplitude calibration at 43 GHz. \\
We compared the gain solutions for different maser lines from each source in our observations, and selected OH 0.55-0.06 (v=1, J=1-0) as the template spectrum source for 43 GHz data, because OH 0.55-0.06 is closer to Sgr A* than VX Sgr (the separation angles of Sgr A* to OH 0.55-0.06 and VX Sgr are $\sim$0$^\circ$.6 and  $\sim$8$^\circ$.5, respectively). Although there are two SiO maser transitions (v=2, J=1-0 and v=1, J=1-0, at 43.122 and 42.820 GHz, respectively), the estimated gain solutions showed little difference. Therefore, we selected the v=1, J=1-0 (42.820 GHz) maser transition since it showed a slightly lower deviation. \\
In order to determine a reference telescope, we have tested several VERA and KVN telescopes, and selected KYS and KUS as the reference telescopes which showed the most stable gain solutions. The reference time ranges for a template spectrum were set at 1 minute when the maser source was observed at the highest elevation (see Table~\ref{tab:obslist}). The effects of template spectrum selection are discussed in detail in Section \ref{sec:discussions}. 

\begin{figure*}[p!]
\centering
\subfloat{\includegraphics[height=100mm, width=\textwidth]{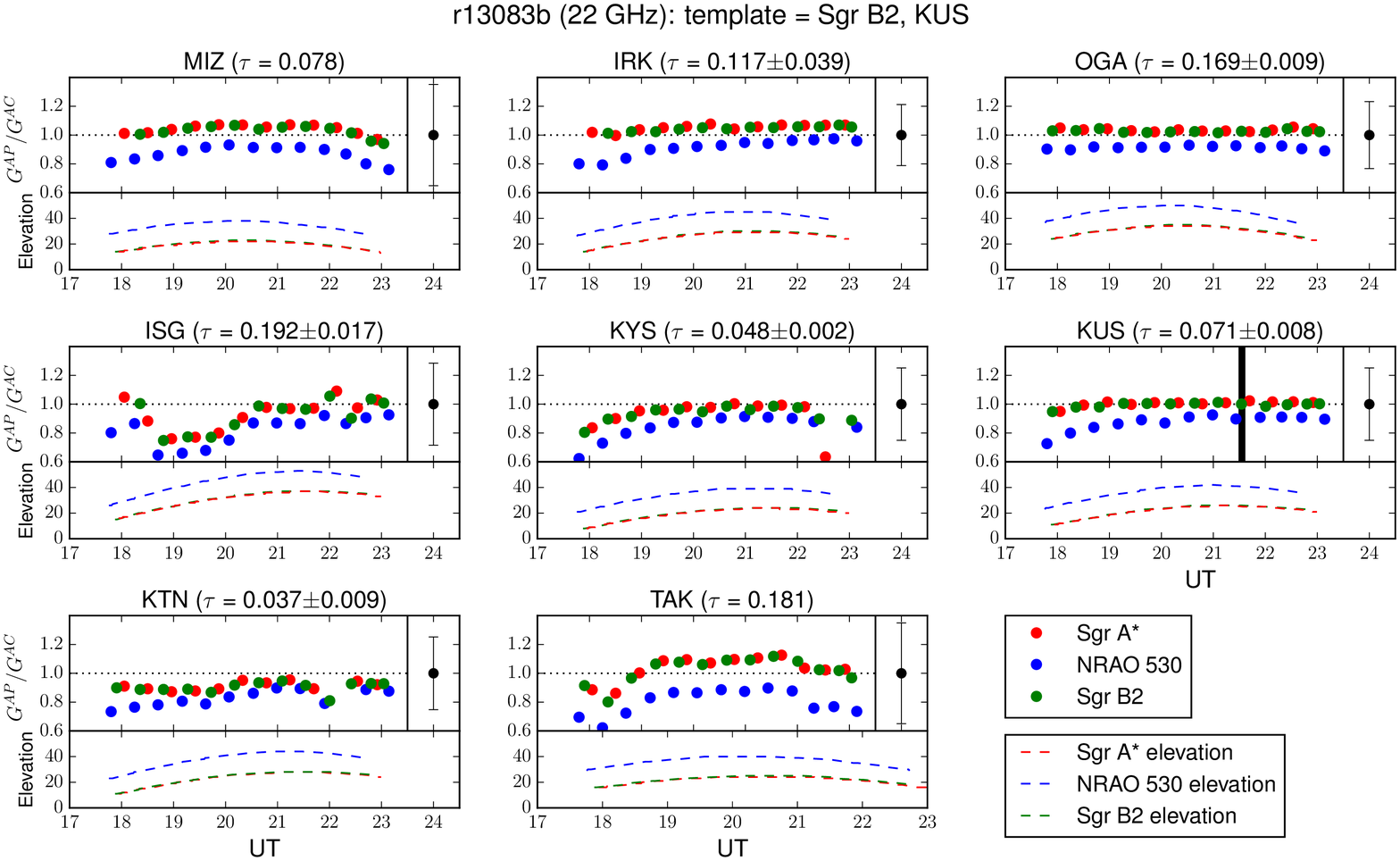}} \\
\subfloat{\includegraphics[height=100mm, width=\textwidth]{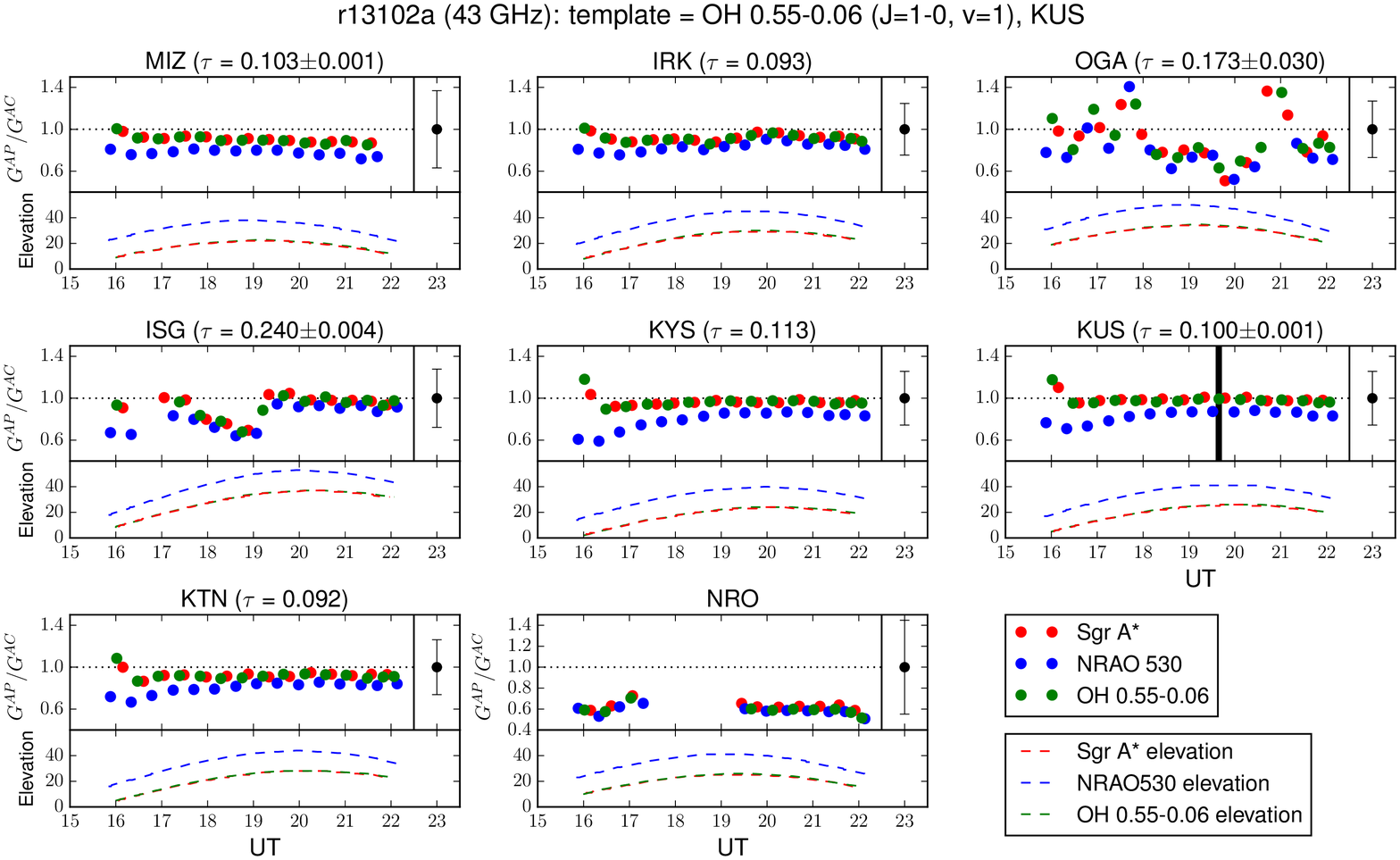}}
\caption{continued.}
\end{figure*}

\begin{figure*}
\ContinuedFloat
\centering
\subfloat{\includegraphics[height=100mm, width=\textwidth]{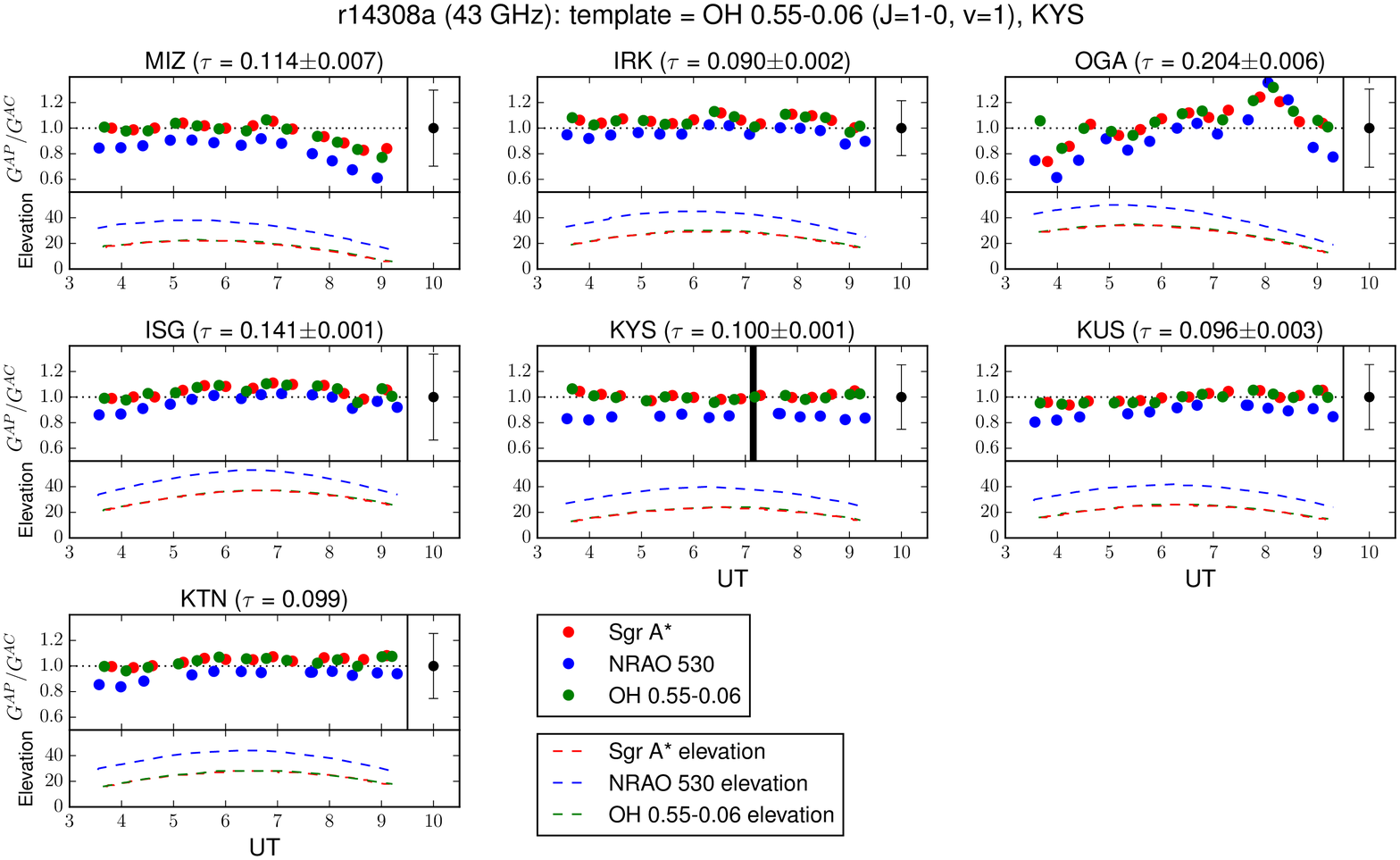}} 
\caption{Gain ratio of the a-priori and template spectrum methods for each source at each telescope (red: Sgr A*, blue: NRAO 530, green: maser sources (i.e. Sgr B2 at 22 GHz and OH 0.55-0.06 at 43 GHz)). The error bars are shown at the right-side of each plot which were estimated by the typical measurement uncertainties (see Appendix A). The vertical thick line (black) shows the time range used as a template spectrum at reference telescope. The broken lines are source elevations for each telescope with time. The a-priori method could be applied to TAK by assuming aperture efficiency $\sim$ 30\% to obtain the gain ratio. Sky opacities to the zenith were measured before and after observations, and are shown next to each telescope's name.\label{fig:gain_ratio}}
\end{figure*}

\subsection{Imaging \label{sec:imaging}}
After phase and gain calibrations, imaging results for NRAO 530 and Sgr A* were obtained using Difmap software (\cite{shepherd:1997}), with UV-data averaged for 30 sec. Deconvolutions were processed using the CLEAN procedure (\cite{hogbom:1974}) and self-calibrations were applied for phase and amplitude. Total flux densities and image dynamic ranges were compared with the naturally weighted map (see Table \ref{tab:image_result}). \\

\section{Results \label{sec:results}}
\subsection{Gain differences \label{sec:gain_differences}}
We compared the gain solutions from the different amplitude calibration methods quantitatively in order to investigate their effects on the amplitude calibration results. Figure \ref{fig:gain_ratio} shows the gain ratio, $G^{AP}/G^{AC}$, for each source at each telescope. For the maser sources and Sgr A*, both methods show good coincidence at all elevations above $\sim$10$^\circ$. NRAO 530 shows larger offsets, because larger gains were obtained from the template spectrum method due to the source separation angle from the maser sources ($\sim$ 15$^\circ$.7 from both Sgr B2 and OH 0.55-0.06). This is discussed further in Section \ref{sec:sep_angle}. Note that the small deviations of gain ratio include the possible pointing offsets as a factor of 1/$\eta_{\theta,i}$, because only template spectrum method corrects the pointing offset as shown in Section~\ref{sec:templatemethod}. The typical values of it are shown in Appendix A. \\
As shown in Figure~\ref{fig:acfitsnr}, ISG in K13, OGA and ISG in Q13, and OGA in Q14 were performed under a bad weather conditions so the gain values from template spectrum method showed lower SNR ($\lesssim$ 3). This results in the large deviations in the gain ratio. \\
Note that the gain ratios in NRO (middle-panel, lower-middle) show that the a-priori method gives $\sim$34\% smaller gain values than the template spectrum method, where the gain solutions from template spectrum method are reliable according to the SNR. It may be attributed to a relatively larger pointing offsets up to $\sim$12 arcsec that can cause $\sim$20\% of flux density loss, as shown in Table~\ref{tab:image_result}. For the time range of UTC 17:30 - 19:20 at NRO, an amplitude drop appeared by a temporal memory flip during the correlation in the VLBI correlation subsystem (VCS) so the data were flagged.

\subsection{Derived gain curve \label{sec:derived_curve}}
\begin{figure}
\centering
\includegraphics[height=60mm, width=83mm]{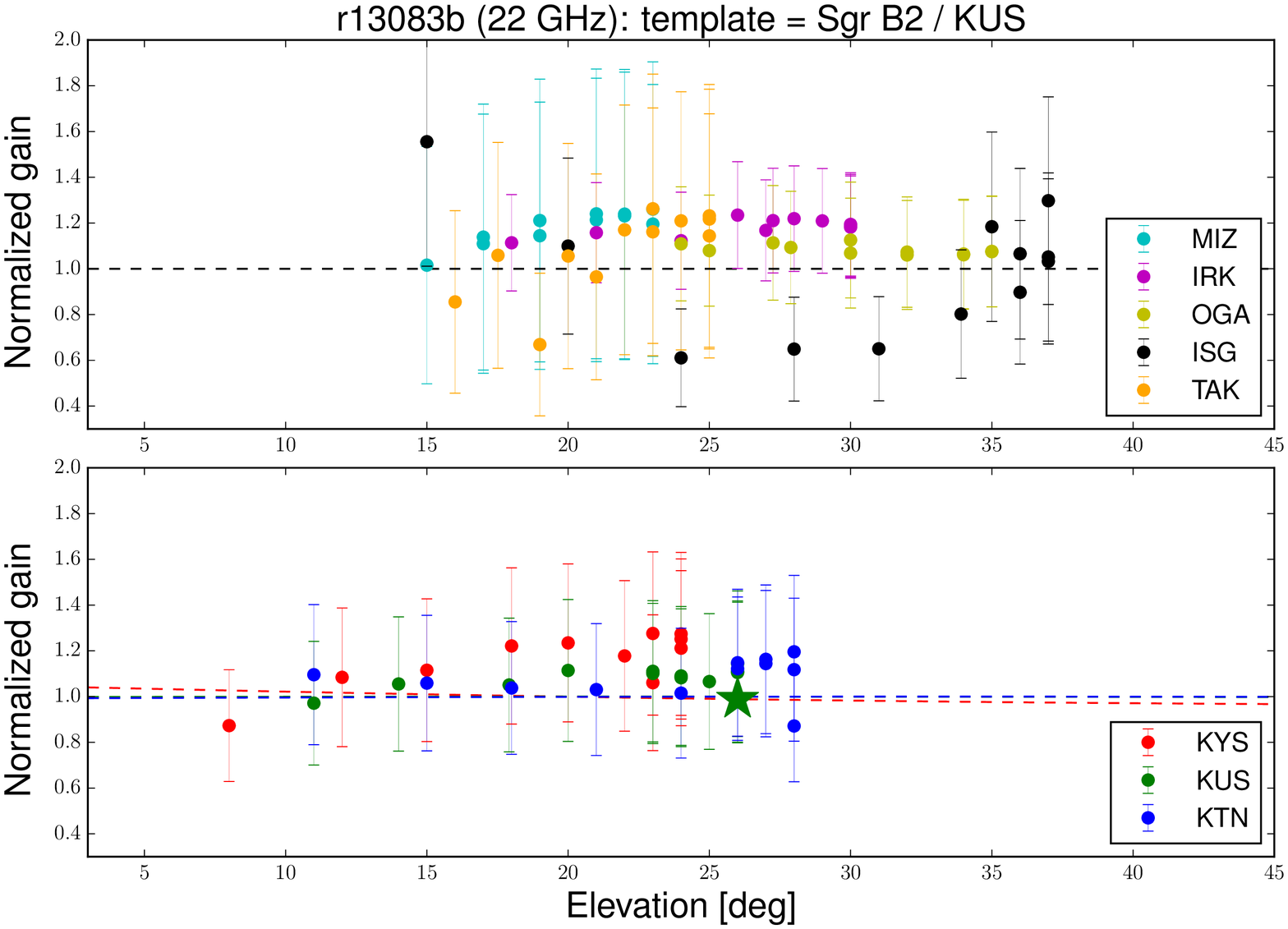} \\
\includegraphics[height=60mm, width=83mm]{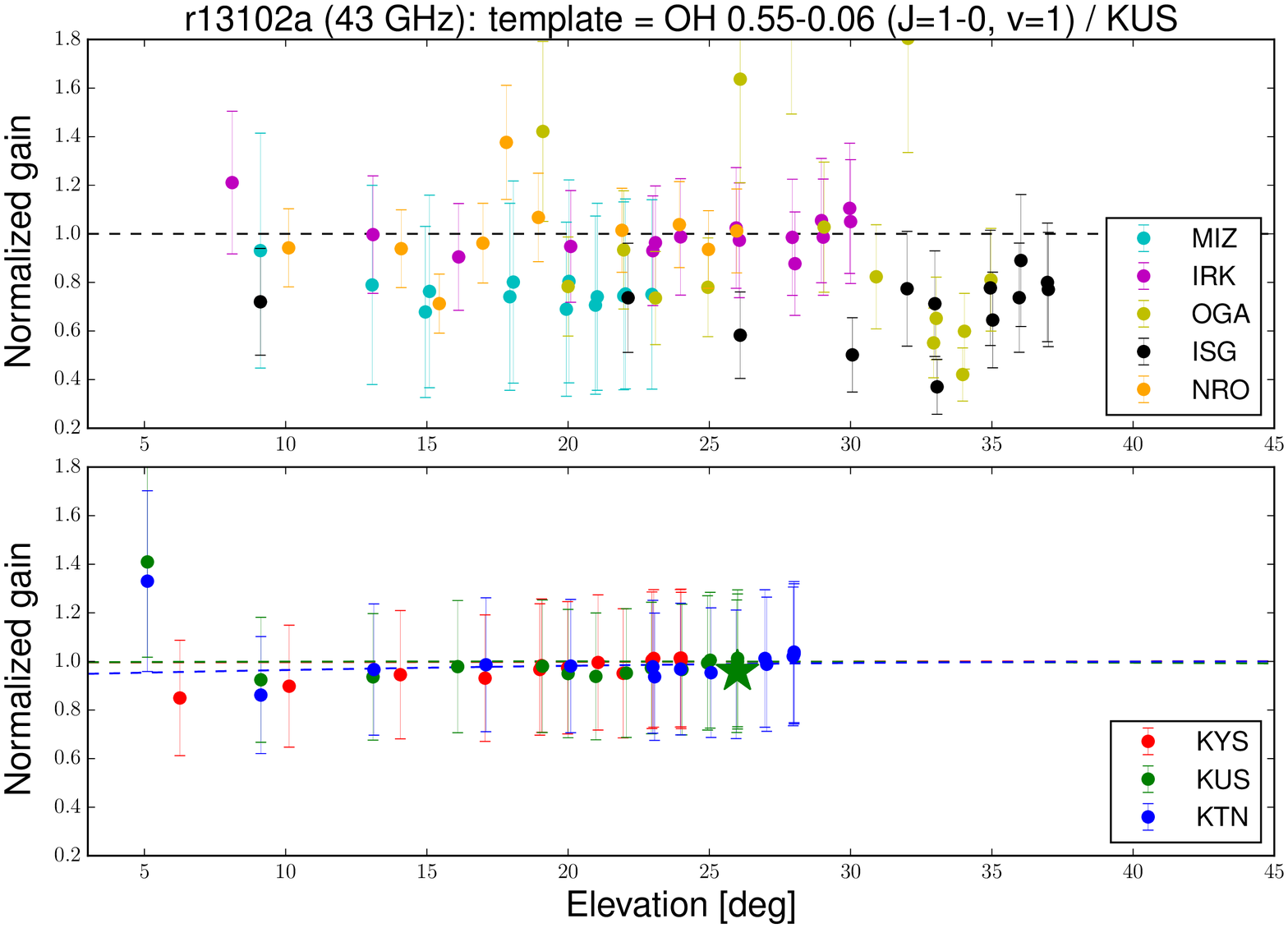} \\
\includegraphics[height=60mm, width=83mm]{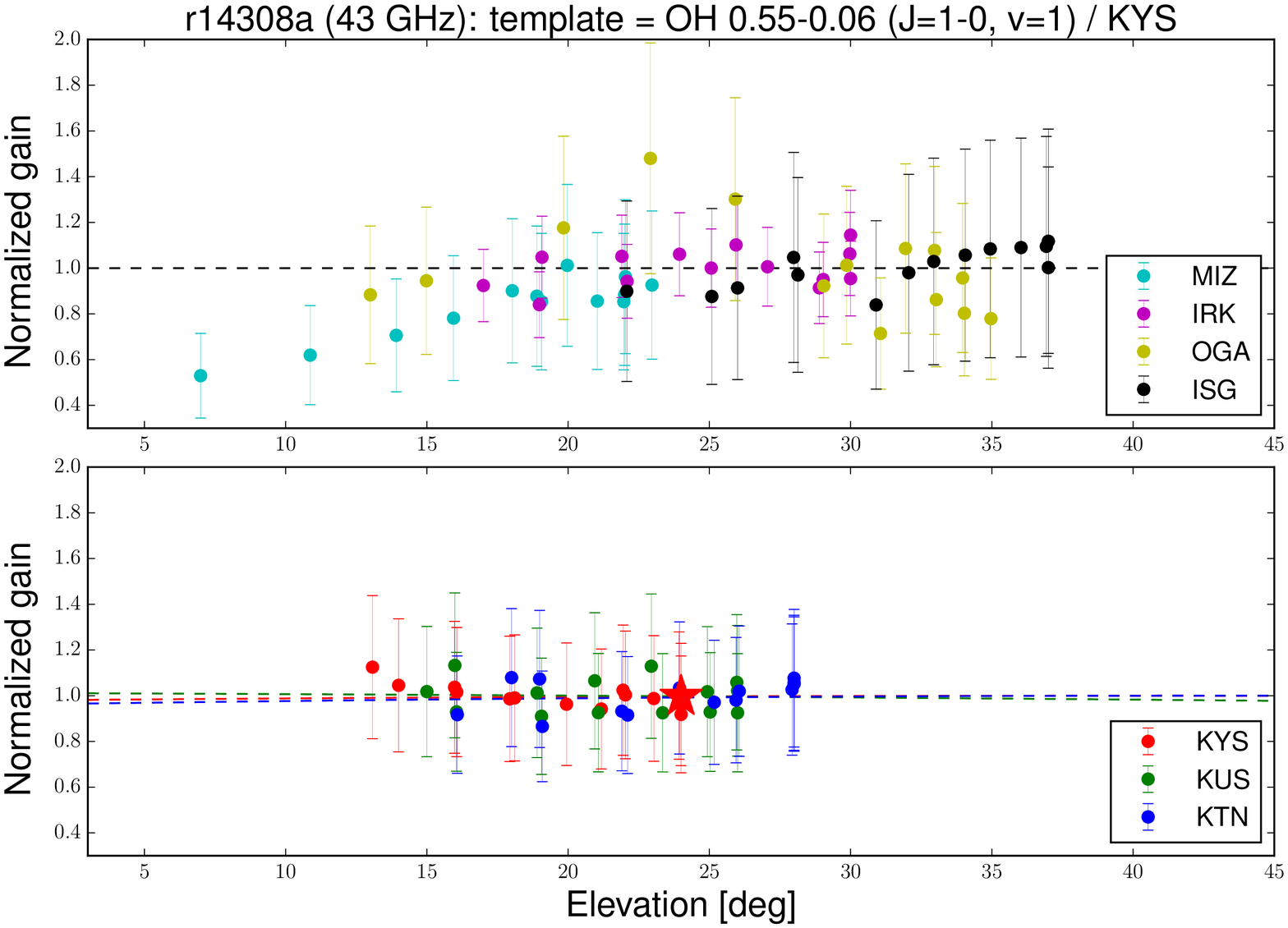}
\caption{Gain curves, derived by the template spectrum method which was normalized with its mean value (i.e. $f_i^{der}(e(t))/ \langle f_i^{der}(e(t))\rangle$; circles), and used in the a-priori method which was normalized with its maximum value ($f_i(e(t))$; broken lines). Each dataset of K13 (top), Q13 (middle) and Q14 (bottom) are shown, and the telescopes in VERA+JVN and KVN are shown in upper and lower side of each panel, respectively. The gains at reference time and telescope are shown with the asterisk mark (KUS in K13, Q13 (green), and KYS in Q14 (red)). The error ranges were estimated by the typical measurement uncertainties (see Appendix A). The difference between two gain curves are $\sim$20\% for $>$ 10$^\circ$ of elevation, while it becomes larger at lower elevations. \label{fig:derived}}
\end{figure}

From the comparison of Equation (\ref{eq:gainsoltru}) and (\ref{eq:gainsol}), a gain curve, $f_i^{der}(e(t))$, can be derived from the gain solutions of template spectrum method as, 
\begin{eqnarray}
\label{eq:derived_curve}
f_i^{der}(e(t)) = \frac{T_{sys, i}^{*}(t)}{\rm DPFU_i} \frac{1}{(G_{i}^{AC}(t))^2}.
\end{eqnarray}
It can be also obtained from the square of the gain ratio (i.e. $(G^{AP}/G^{AC})^2$) because it corresponds to a gain curve ratio, $f_i^{der}(e(t))/f_i(e(t))$. \\
Figure \ref{fig:derived} shows the derived gain curves normalized to the mean value at each telescope, and gain curves used in the a-priori method for VERA, JVN and KVN. The difference between two gain curves is about 20\% for elevations $>$10$^\circ$, except for ISG in K13, ISG and OGA in Q13, and OGA in Q14 where the gain SNRs of template spectrum method are lower than $\sim$3. Note that the gain curve is unity at all elevations for VERA, whereas KVN uses a polynomial function based on cross-scan observations from 20$^\circ$ to 80$^\circ$ elevation. \\
Based on typical pointing accuracy of each telescope in KaVA, the pointing offset of $\sim$5 arcsec may cause a gain loss of $\sim$2 \%. Other factors, such as $T_{sys}^{*}$ and ${\rm DPFU}$ measurement uncertainties, are responsible for the 5-15\% of losses. On the other hand, the additional losses are expected for low elevation observations ($<$ 10$^\circ$) due to atmospheric opacity changes. Therefore, it is strongly recommended to remove data from elevation 10$^\circ$. In practice, KaVA observing schedules have been optimized to observe sources above $\sim$10$^\circ$ since 2014 except for the MIZ where is located at the most east-north side among the KaVA stations, so both of amplitude calibration methods are applicable within 10\% difference.

\subsection{Amplitude calibrations for the additional telescopes (TAK, NRO) and its effects on the image sensitivity \label{sec:DR}}
The template spectrum method has the advantage to extend possible telescopes comprising VLBI. While the a-priori amplitude calibration method requires the gain curve and $T_{sys}^{*}$ measurements for all telescopes, the template spectrum method needs only those values for a template spectrum (i.e. at one station at a short time range). Therefore, any radio telescopes without $T_{sys}^{*}$ and gain curve information could still participate in VLBI observations using this method for their amplitude calibrations. \\
In our observations, for example, TAK and NRO participated in addition to KaVA in K13 and Q13, respectively. While both telescopes have the $T_{sys}^{*}$ measurements, the TAK gain curve information was assumed as 30\%, and NRO was suspected to have a relatively larger pointing offsets. Therefore, the imaging results of a-priori method had to be compared with the results from template spectrum method which can give a more practical gain curve without pointing offset effects (see Table~\ref{tab:image_result}). \\
We compared the total flux densities from each method by applying 1) phase-only self-calibration and 2) phase+amplitude self-calibration, to see the effects of gain solutions on the resultant flux densities more strictly. As a result, the measured total flux densities from each method were consistent within $\sim$10\% for both 1) and 2) cases, except for the KaVA and KaVA+NRO in Q13 where the NRO had a severe gain loss. Note that NRAO 530 flux densities from the template spectrum method were corrected by the separation angle correction factor  ($\beta^2$ in Table \ref{tab:image_result}, see also Section \ref{sec:sep_angle}). The total flux densities of NRAO 530 were compared with relevant imaging results from archival data in Boston University (BU) Blazar Group, and the results from both methods were well consistent with the BU archival data. \\
In addition, the participations of TAK and NRO gave a notable dynamic range increase in the resultant images about 10-15\%. This implies that more radio telescopes can compromise the VLBI using the template spectrum method which were not available to calibrate their amplitudes before so that it can provide better sensitivity. Therefore, the template spectrum method is crucial to extend the candidate radio telescopes for EAVN observations. 

\section{Discussions \label{sec:discussions}}
Gain uncertainties of the a-priori method arise from inaccuracies in $T_{sys}^{*}$ measurements and gain curves. On the other hand, the template spectrum method derives a gain curve from the $\rho$-ratio during the observation time, so could provide more practical gain values. Under ideal observation with a single maser line from a compact source without pointing errors, the ``derived" gain curve will trace the practical changes of aperture efficiency during observations better than a-priori method as long as it holds the fundamental assumptions: no variations of a maser line during the observation. However, the template spectrum method uncertainties arise from reference selection, maser line profile, spatial distributions of maser components and separation angle from a target. 

\begin{figure*}
\centering
\includegraphics[height=65mm, width=\textwidth]{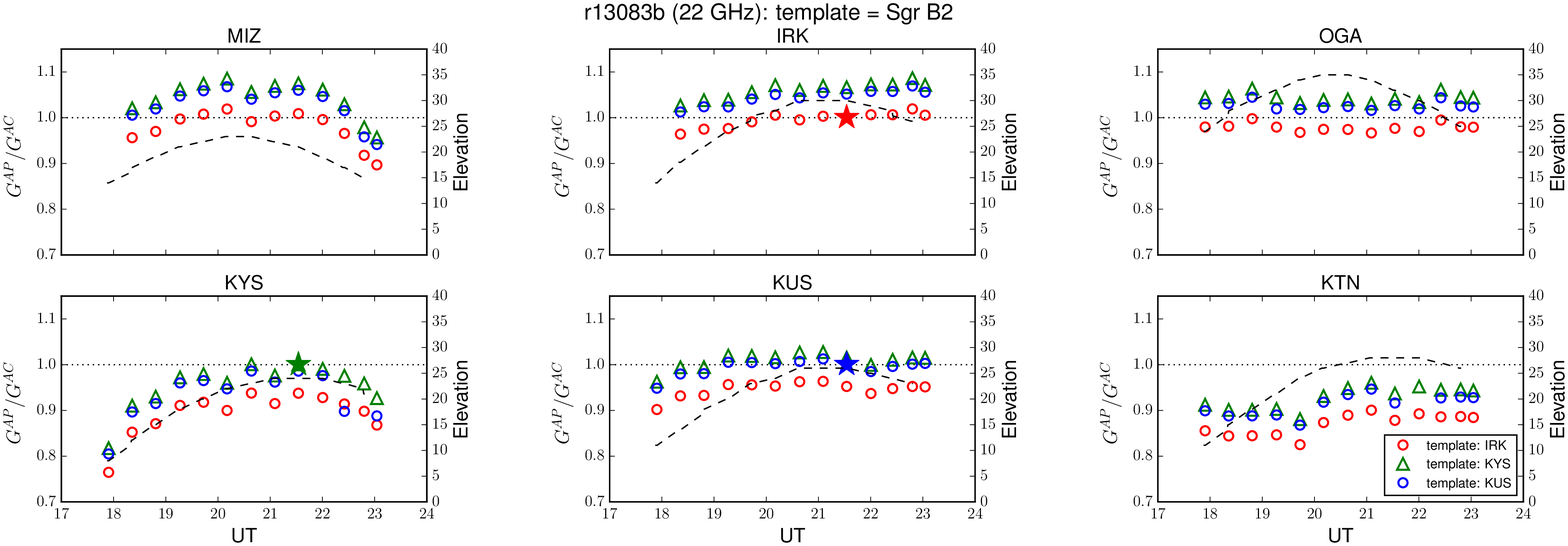}
\includegraphics[height=65mm, width=\textwidth]{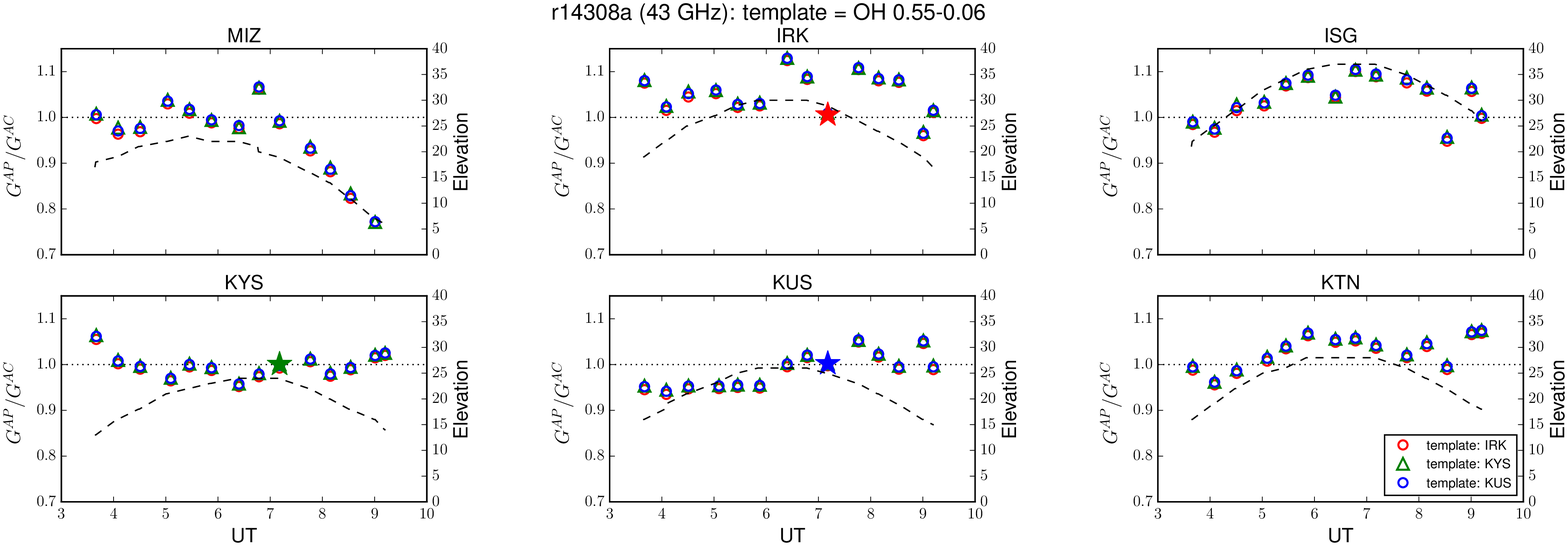}
\caption{The effect of reference telescope selection in the template spectrum method. Each color represents the reference telescopes (red: IRK, green: KYS, blue: KUS). The maser source used for the template spectrum method is same as in Figure \ref{fig:gain_ratio}. The broken line is the maser source elevation at each telescope. The gain ratios at reference time and telescope are shown with the asterisk mark.
\label{fig:temp_tel}}
\end{figure*}

\subsection{Template spectrum selection \label{sec:template_telescope}}
When selecting a template spectrum, the following criteria are important: 1) select a reference telescope that shows the clearest maser line total-power spectrum without any ambiguous peaks or fluctuating baseline, and 2) select a reference scan when the source is at elevation where aperture efficiency of the reference telescope is maximized. \\
From Equation (\ref{eq:gainsol}), gain solutions from the template spectrum method depend on $\rho T_{sys}^{*} / f(e) \rm DPFU$ of the template spectrum. Since $\rho$ is proportional to $A_{e}/T_{sys}^{*}$, this implies the effective aperture size of the template spectrum relative to the zenith (i.e. $A_{e}$(template)/$A_{e}(e = 90^\circ$)). Note that all $A_{e}$ for different times and telescopes are normalized with the effective aperture size of the reference telescope at reference scan, which corresponds to the derived gain curves. Therefore, the reference scan and telescope should be determined as having the most stable gain and $T_{sys}^{*}$ values where usually the largest effective aperture area is given during the observing time. In addition, to minimize $\rho$-measurement error, the template spectrum is strongly recommended to have a single-sharp maser line feature with a flatter baseline in its total-power spectrum. Note that the different bandpass responses at each telescope can give different spectral baselines so the bandpass calibration has to be carefully applied first. 
\\
Figure~\ref{fig:temp_tel} shows the gain ratio shifts for the different reference telescopes as IRK in VERA, KYS and KUS in KVN. Reference time ranges were considered when the maser source elevation was highest ($\sim$ 30$^\circ$), near where KVN telescope aperture efficiencies are normally maximized ($\sim$48$^\circ$). Consequently, gain results from the template spectrum method are reasonably close to the a-priori method results (i.e. gain ratio $\approx$ 1), particularly at 43 GHz in the bottom panel of Figure \ref{fig:temp_tel}. \\
However, at 22 GHz, it was difficult to find which telescope provides the closest gain to a-priori method results. For example, the MIZ gain ratio was close to unity when IRK was selected as the reference telescope, but there was larger offset for KTN. It is clearly seen only at 22 GHz data, so an additional effect from the maser source itself is expected and is discussed in the following section.

\begin{figure*}
\centering
\includegraphics[height=58mm, width=75mm]{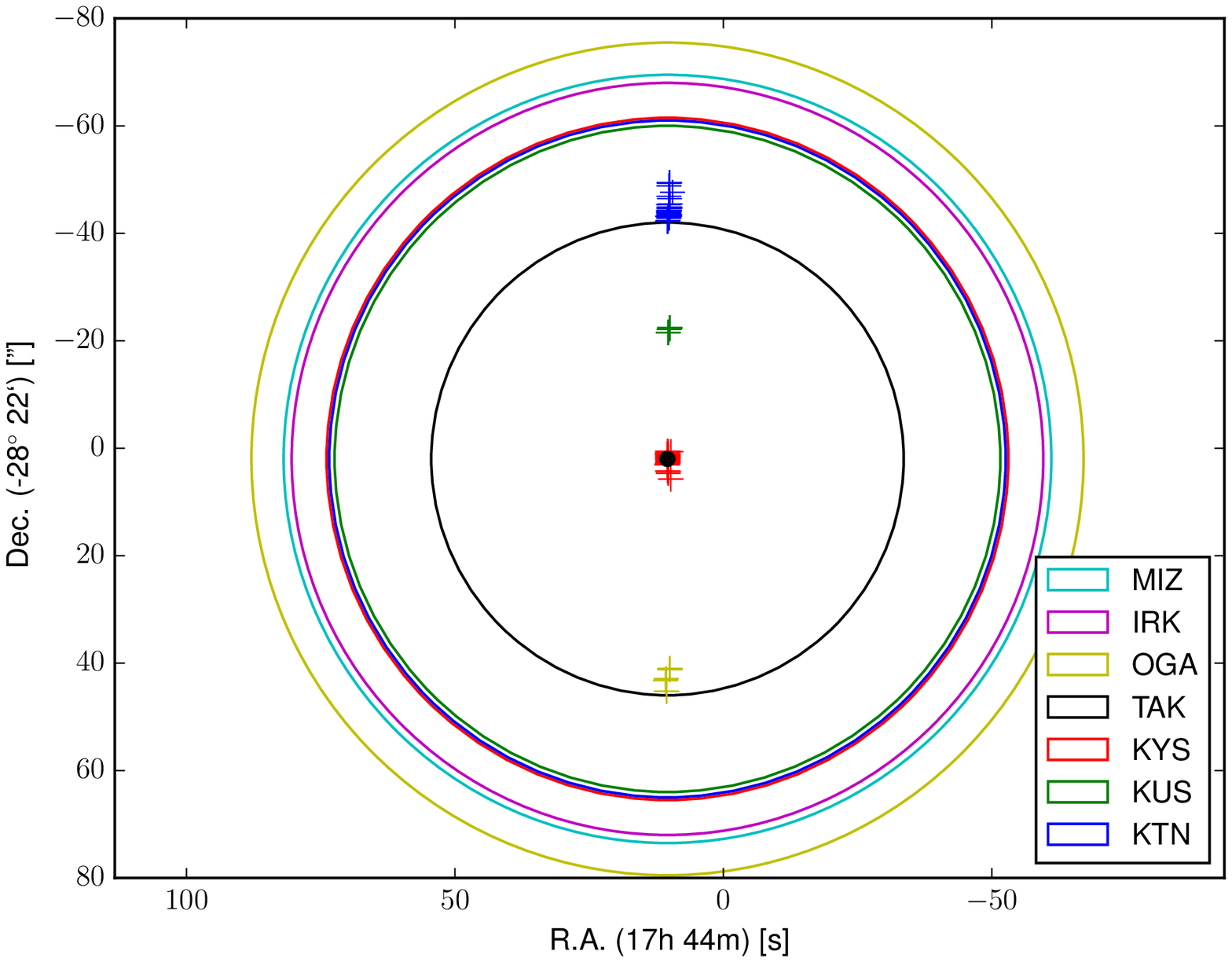}
\includegraphics[height=60mm, width=77mm]{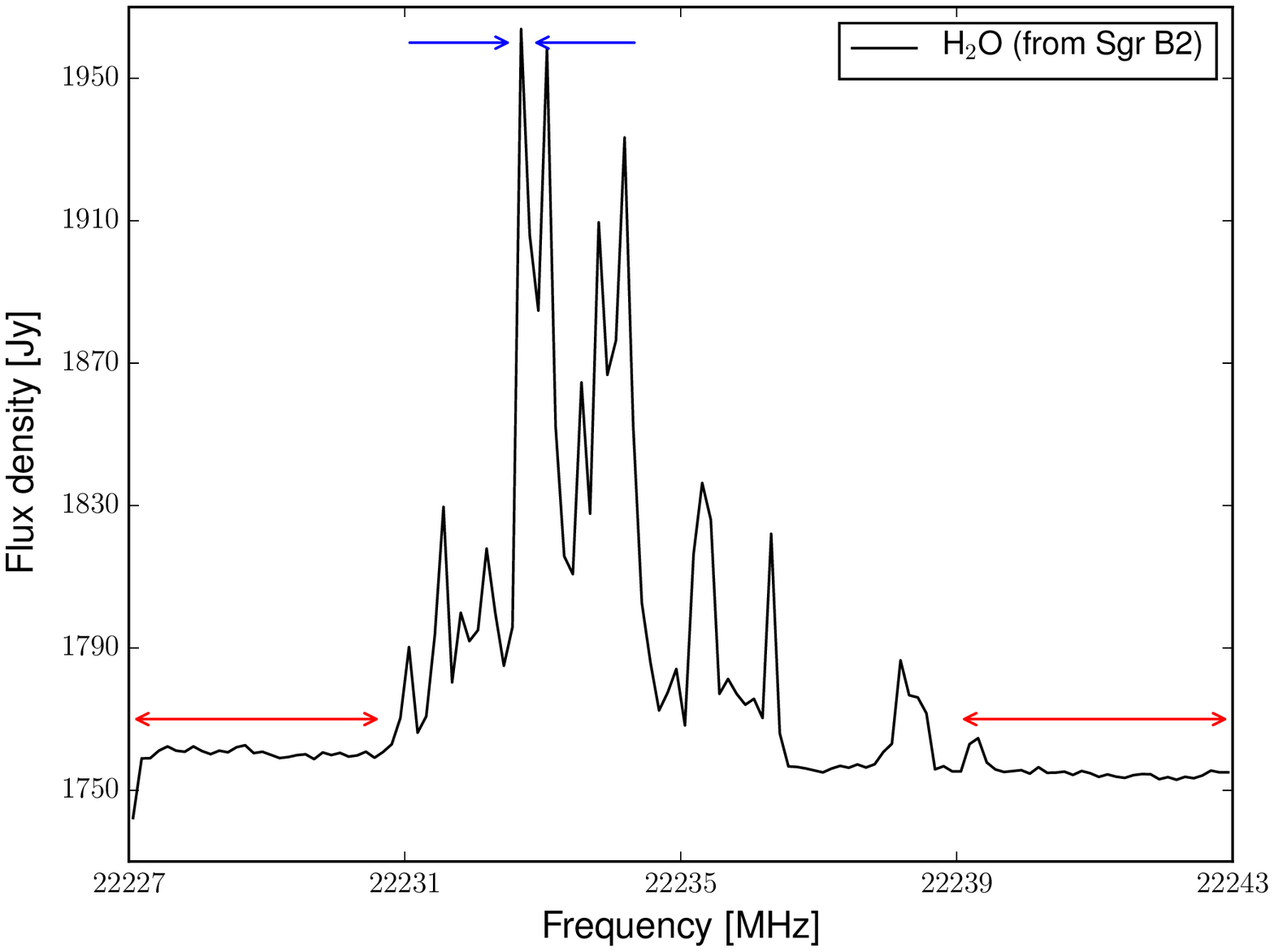}
\caption{(Left) The H$_{2}$O maser emission spots in Sgr B2, referred to \citet{mcgrath:2004} (colored crosses: red (main), blue (north), green (mid-north) and yellow (south)). Colored circles represent half-power beam-width (HPBW) of each telescope, pointing to the center of the regions (black point). (Right) The total-power spectra of H$_2$O maser line in Sgr B2 observed in K13, with bandpass calibration using NRAO 530. Note that the flux densities were obtained through a-priori amplitude calibration. We selected a peak at 22232.7 MHz (blue-arrows) and the baseline ranges around 22228 and 22242 MHz (red-arrows). However, multiple maser peaks are evident, and contributions from each emission regions are mixed. Therefore, if a pointing shifted from its center during an observation, the relative powers of each peak differ which makes it difficult to measure accurate elevation dependent gain changes. 
\label{fig:sgrb2}}
\end{figure*}

\subsection{Maser source selection for the template spectrum method \label{sec:maser_source}}
Since the template spectrum method uses total-power spectrum of a maser line, it is important to constrain a reliable frequency range corresponding to the peak and baseline to evaluate the $\rho$ values. In addition, the maser source spatial distribution should be compact enough compared to the telescope beam size. Otherwise, the beam size differences convolved with telescope pointing errors may introduce additional gain differences. \\
For example, in K13, the H$_2$O maser of Sgr B2 was used for the template spectrum method. Sgr B2 has a largely extended molecular cloud up to 120 arcsec in the declination direction from the pointing center, and the H$_2$O maser distributions in each region (named as, main, north, mid-north and south) are also extended having similar velocities (\cite{mcgrath:2004}; Figure \ref{fig:sgrb2} (Left)). Since all KaVA telescopes has a beam size $>$ 120 arcsec at 22 GHz and different effective aperture sizes, it is difficult to separate the amount of contributions from each emission regions in Sgr B2 to each telescope, even if the maser line peak frequency range is set as narrow as possible to avoid ambiguity (Figure \ref{fig:sgrb2} (Right)). This makes gain solutions from the template spectrum method more sensitive to the frequency range selection where the maser line shows a peak, as well as reference telescope selection. \\  
On the other hand, SiO masers usually show one sharp peak in a very narrow frequency range (see, for example, Figure \ref{fig:rho}) and sufficiently compact, because its emitting region is spatially limited near an evolved star. Therefore, SiO masers are generally good candidates for this application. 

\begin{table*}
\tbl{The imaging results.}{%
\centering
\begin{tabular}{lllcccc}
\hline
NRAO 530 & ($\beta^2 \approx$ 0.74$\pm$0.09)$^{\rm a}$ &&&& \\
Date  &  Frequency  &  Method$^{\rm b}$ / Array  & Total flux density$^{\rm c}$ (Jy) &   $\beta^2$-corrected flux density (Jy) & Dynamic range  \\
\hline
2013.03.28  &  22 GHz  & AP / KaVA  	  & 3.96 (3.93)  & - &   1512    \\
         	   &  22 GHz  & AP / KaVA+TAK  & 3.93 (3.87)  & - &   1737    \\
      		   &  22 GHz  & TS / KaVA  	  & 4.98 (4.98)  & 3.69 $\pm$ 0.45 &  1391  \\
		   &  22 GHz  & TS / KaVA+TAK  & 5.06 (5.04)  & 3.73 $\pm$ 0.45 &  1616  \\
\hline
2013.04.12  &  43 GHz  & AP / KaVA 	   & 3.02 (2.78)  & - &  1141  \\
       		   &  43 GHz  & AP / KaVA+NRO  & 3.09 (2.39$^{\rm *}$)   & - &  1253  \\
		   &  43 GHz  & TS / KaVA 	   & 4.08 (3.76)   & 2.78 $\pm$ 0.34 &    1012  \\
		   &  43 GHz  & TS / KaVA+NRO  & 3.84 (3.57)   & 2.64 $\pm$ 0.32 &  1112  \\
2013.04.17  &  43 GHz  & AP / VLBA (BU)   &  2.73  & - &  -  \\
\hline
2014.11.04  &  43 GHz  &  AP / KaVA  & 2.25 (2.23)  & - &      662   \\
 		   &  43 GHz  &  TS / KaVA  & 2.81 (2.77)     & 2.05 $\pm$ 0.25 &   667  \\
2014.11.15  &  43 GHz  &  AP / VLBA (BU) &  2.10  & - & -   \\
\hline
\hline
Sgr A* & ($\beta^2 \approx$ 1.00)$^{\rm a}$ &&&&& \\
Date  &  Frequency  &  Method$^{\rm b}$ / Array  & Total flux density$^{\rm c}$ (Jy) &&  Dynamic range  \\
\hline
2013.03.28  &  22 GHz  & AP / KaVA  	  & 1.03 (0.97)  &&   560    \\
    		   &  22 GHz  & AP / KaVA+TAK  & 1.04 (0.95)  &&   640    \\
		   &  22 GHz  & TS / KaVA  	  & 1.08 (1.01)  &&  568  \\
	   	   &  22 GHz  & TS / KaVA+TAK  & 1.05 (0.99)  &&  638  \\
\hline
2013.04.12  &  43 GHz  & AP / KaVA  	   & 1.12 (1.02)   &&  795  \\
     		   &  43 GHz  & AP / KaVA+NRO  & 1.15 (0.84$^{\rm *}$)  &&  900  \\
		   &  43 GHz  & TS / KaVA 	   & 1.23 (1.14)   &&    821  \\
		   &  43 GHz  & TS / KaVA+NRO  & 1.24 (1.16)   &&  945  \\
\hline
2014.11.04  &  43 GHz  &  AP / KaVA  & 1.58 (1.47)  &&      1049   \\
 		   &  43 GHz  &  TS / KaVA  & 1.49 (1.40)     &&   1056  \\
\hline
\end{tabular}}\label{tab:image_result}
\begin{tabnote}
Total flux densities and image dynamic ranges from each method: a-priori (AP) and template spectrum (TS). The images were naturally weighted. The total flux densities have been compared to the adjacent observations in Boston University (BU) data for NRAO 530. \\
$^{\rm a}$ The separation angle correction factor (see Section~\ref{sec:sep_angle}). \\
$^{\rm b}$ The results of template spectrum method were obtained using OH 0.55-0.06 (v=1, J=1-0). \\
$^{\rm c}$ The quantization loss correction factors of 1.35 and 1.3 were multiplied to each data in 2013 and 2014, respectively (\cite{lee:2015b}). Note that each values without/with parenthesis resulted from the image reconstruction through the iterative processes of CLEAN and self-calibration for phase only, and applying the amplitude self-calibration onto the former results, respectively. \\
$^{\rm *}$ The flux density loss seems to be dominated by the pointing offset ($<$12 arcsec) at NRO. \\
\end{tabnote}
\end{table*}
%
\subsection{Separation angle effect \label{sec:sep_angle}}
\begin{figure*}
\centering
\includegraphics[height=45mm, width=0.325\linewidth]{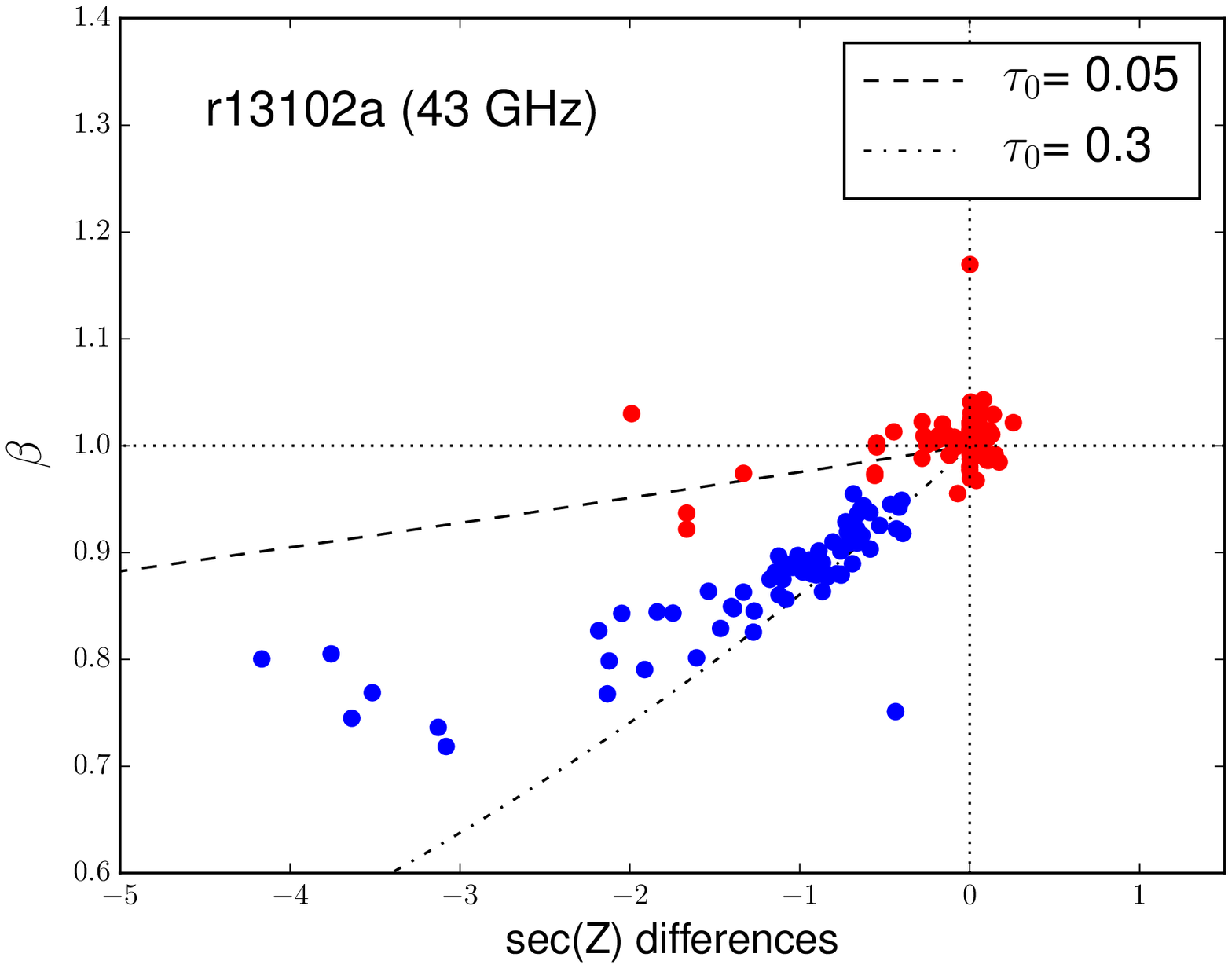}
\includegraphics[height=45mm, width=0.325\linewidth]{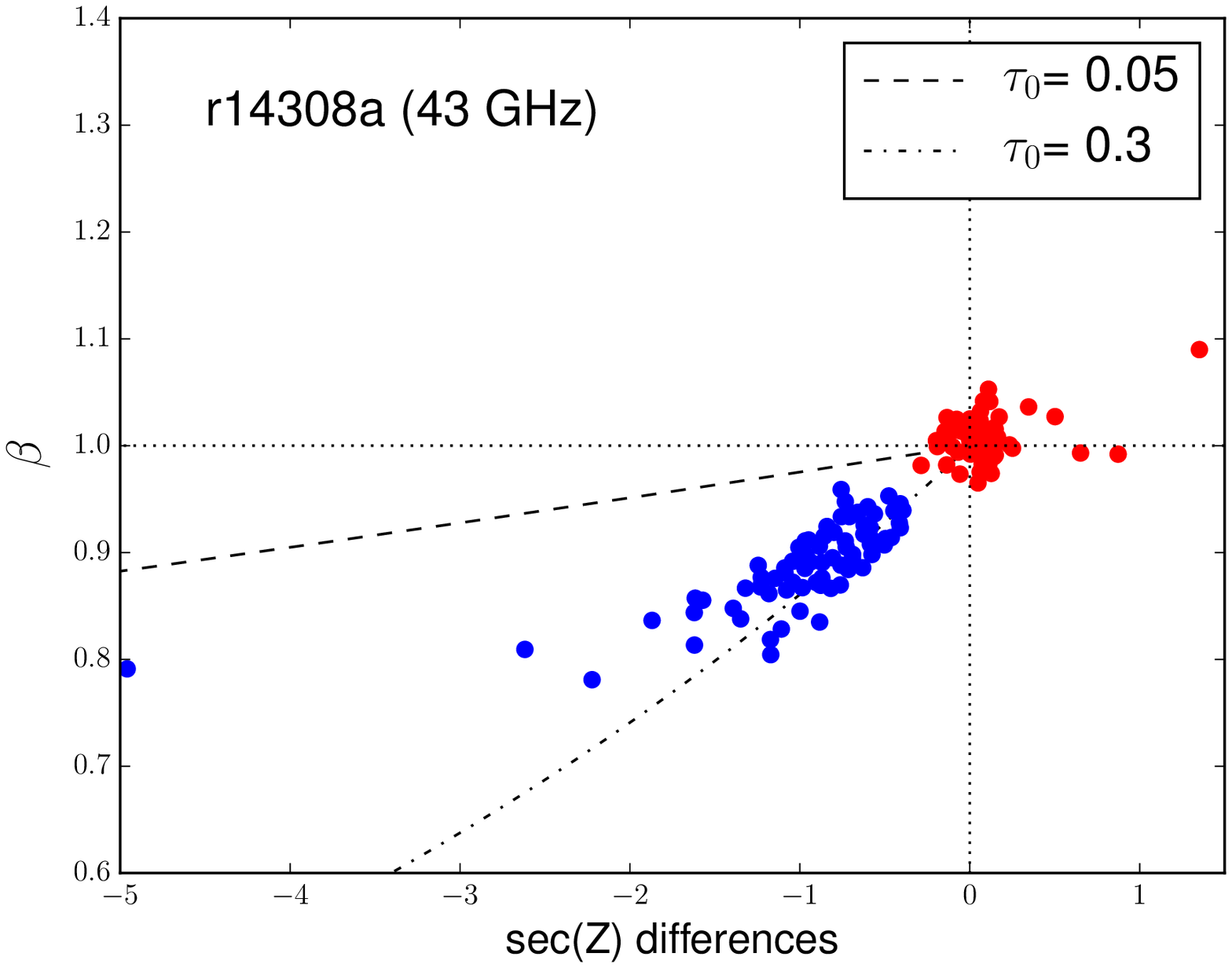} 
\includegraphics[height=45mm, width=0.325\linewidth]{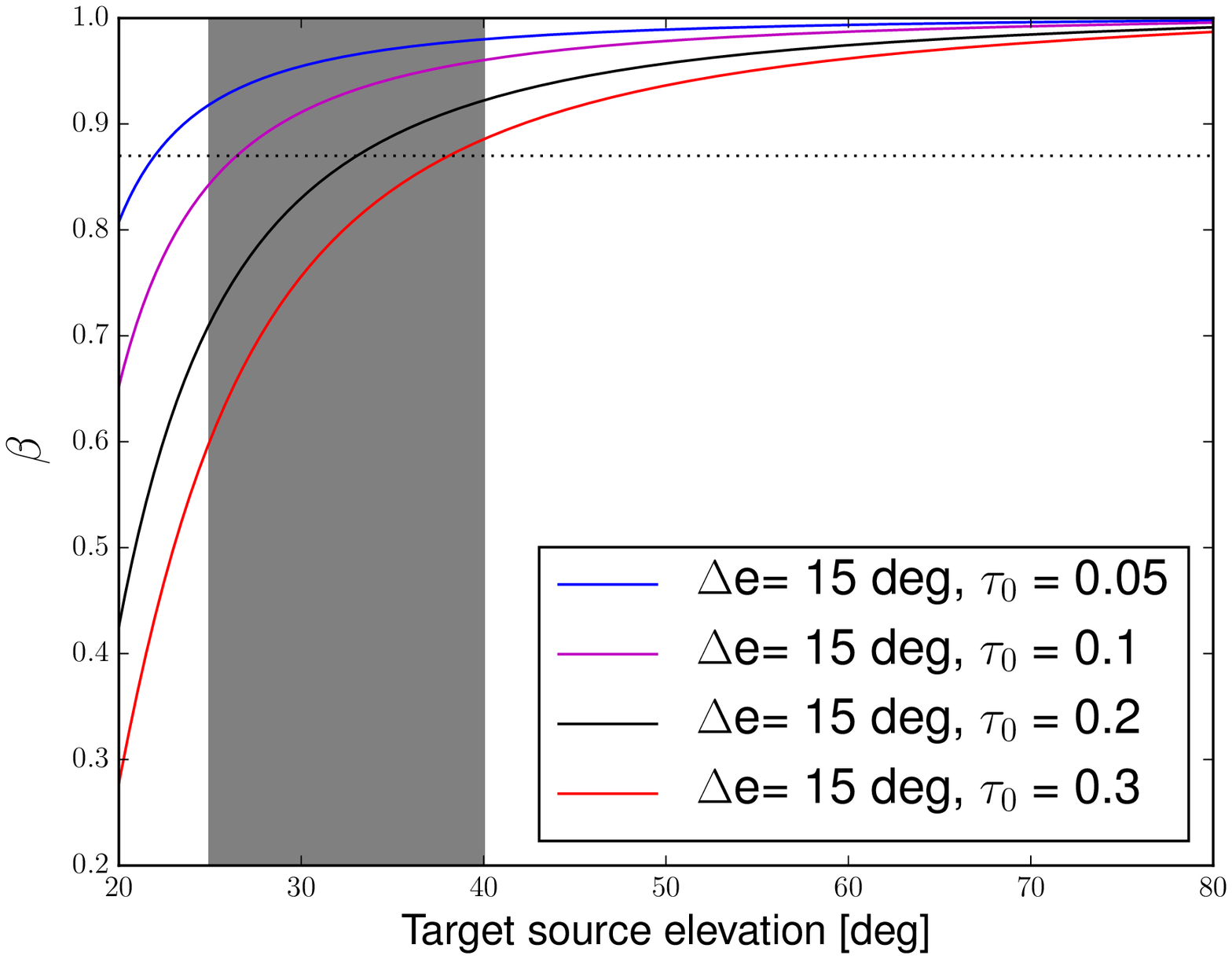} 
\caption{(Left and Middle) Observational estimate of $\beta$ for Sgr A* (red-circle) and NRAO 530 (blue-circle) depending on the sec(Z) differences in each observation of Q13 (left) and Q14 (middle), where Z is the zenith angle. Note that OGA in Q14 was excluded by the bad weather. Each line represents the estimated $\beta$ under the minimum and maximum zenith-opacity (dashed line: 0.05 and dash-dotted line: 0.3).
(Right) The estimated $\beta$ along with the target source's elevation, 15$^\circ$ away from its maser source, under the different opacity to the zenith. The horizontal dotted-line represents the $\beta$ = 0.87 which is the averaged value within the elevation range of NRAO 530 in our observations (shaded region in right panel) where $\tau_0$ = 0.05 $\sim$ 0.3.
\label{fig:beta_elev}}
\end{figure*}

\begin{figure*}
\centering
\includegraphics[height=45mm, width=80mm]{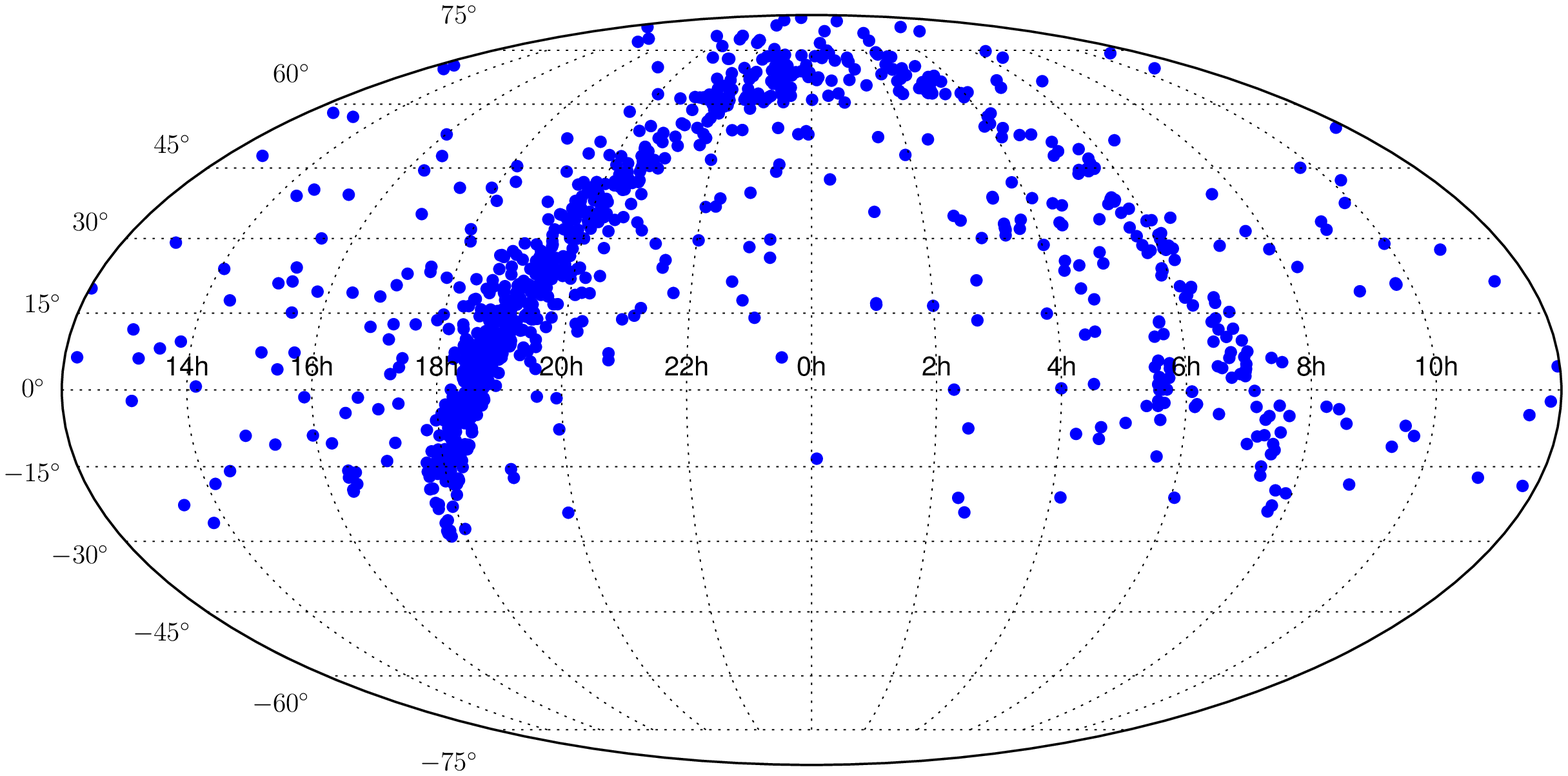} 
\includegraphics[height=45mm, width=80mm]{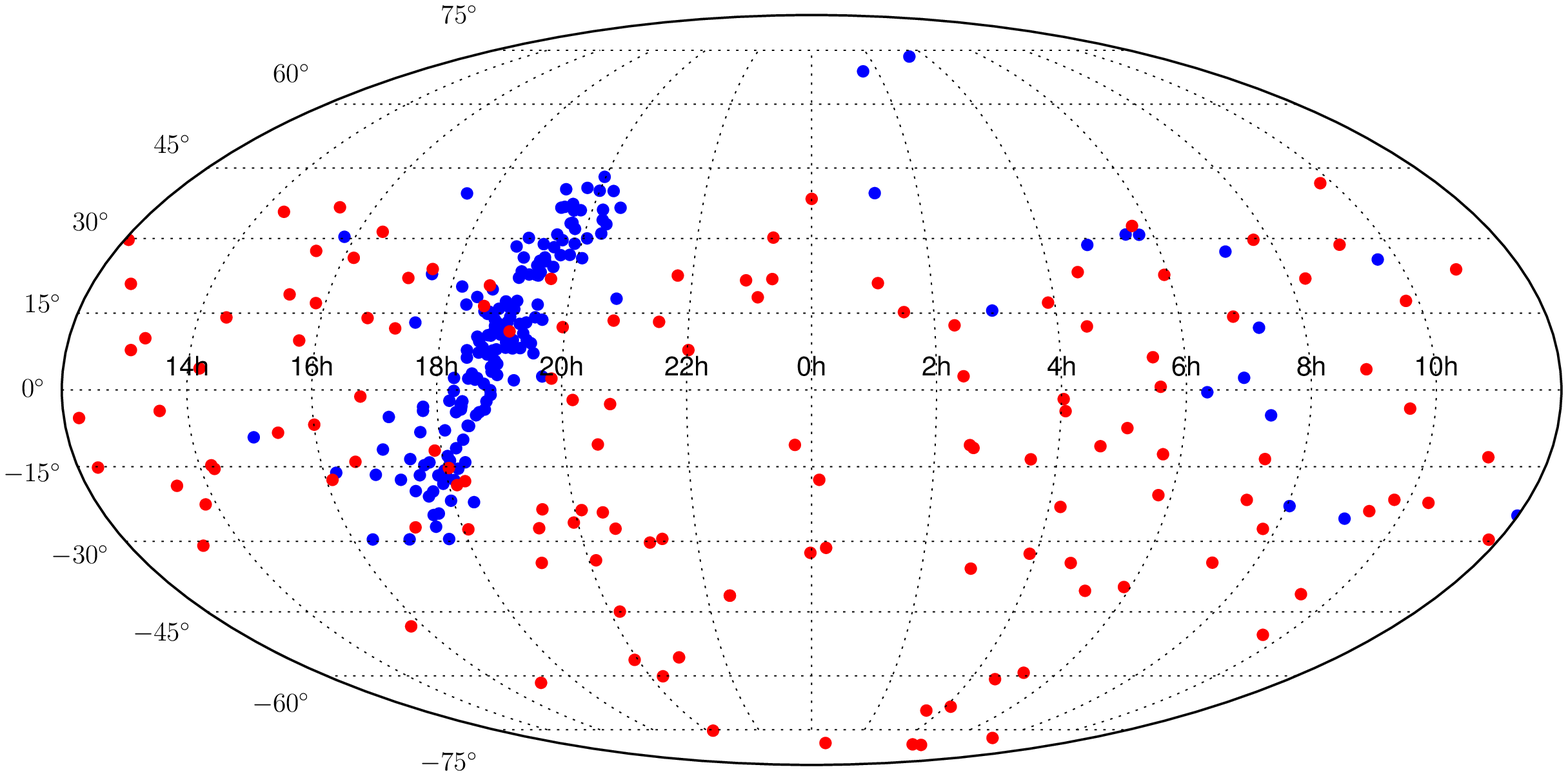} 
\caption{The sky distribution of H$_2$O maser (left: \cite{valdettaro:2001}) and SiO maser (right: \cite{indermuehle:2013, yoon:2014}; Cho et al. in prep.) sources with the number of 1013 and 306, respectively. Note that the H$_2$O maser (left, blue-circles) and a part of SiO maser (right, blue-circles) source surveys were carried out using Arcetri observatory in Italy and KVN, respectively, where are located in the northern hemisphere so that only sources of Dec $\gtrsim$ -30$^\circ$ were available. The SiO maser source survey for lower declinations was carried out using the Mopra telescope in Australia in \citet{indermuehle:2013} (right, red-circles).  \label{fig:mas_dist}}
\end{figure*}

Figure \ref{fig:gain_ratio} shows that the gain ratio between the a-priori and template spectrum methods for NRAO 530 has a larger offset from unity compared to Sgr A*. The gain solutions from template spectrum method were obtained from a nearby maser source and applied to the target source by interpolation along time, so the angular distance between two sources is often ignored. \\
In our observations, maser sources were found within 1$^\circ$ from Sgr A* whereas the separation to NRAO 530 was $\sim$15$^\circ$. Since maser source elevations were lower than NRAO 530, the interpolated gain values for NRAO 530 from masers (Sgr B2 and OH 0.55-0.06) were higher than the gain from the a-priori method about 20$\sim$30\% on average, so this produced the corresponding flux density increases. Therefore, interpolation errors introduced by the separation angle should be corrected when the template spectrum method is applied to distant source from a maser source. \\
Let the gain ratio of the maser source itself be ${G^{AP}(t')}/{G^{AC}(t')} \equiv \alpha(e(t'))/\sqrt{\eta_{\theta}(t')}$, where $t'$ is the observing time for the maser source, $e(t')$ is the elevation and $\eta_{\theta}(t')$ is gain efficiency when a pointing offset, $\theta$, is given. Then the separation angle correction factor, $\beta(\Delta e(t))$, can be obtained from the gain ratio for a target source as,
\begin{eqnarray}
\label{eq:ampratio}
\beta(\Delta e(t)) \approx \frac{G^{AP}(t)}{G^{AC}(t)} \frac{1}{\langle \alpha(e(t')>10^\circ)\rangle} 
    \frac{\langle \sqrt{\eta_{\theta}(t')} \rangle}{\sqrt{\eta_{\theta}(t)}} ,
\end{eqnarray} \\
where $t$ is the target observing time and $\langle \alpha(e(t')>10^\circ) \rangle$ is averaged gain ratio of the maser source above 10$^\circ$ in elevation to minimize errors, which is $\approx$ 1. The pointing offset effects are represented with ($\eta_{\theta}$(t$'$)/$\eta_{\theta}$(t))$^{1/2}$, as explained in Section~\ref{sec:gain_differences}. This can be approximated as unity if no significant pointing changes are shown between target and maser sources. $\beta(\Delta e(t))$ represents the interpolation errors by the separation angle where $\Delta e(t)$ is an elevation difference between the maser and target sources at an observing time, $t$. Note that the gain ratios at time $t$, $G^{AP}(t)/G^{AC}(t)$, for NRAO 530 and Sgr A* are shown in Figure~\ref{fig:gain_ratio}. Therefore, the gain values for sources distant from a maser source can be reasonably estimated by applying $\beta(t)$.  \\
In our measurements, elevation differences between sources were almost constant during the observing time. Therefore, the correction factors were assumed to be constant (i.e. $\beta(\Delta e(t)) \approx \beta$), and $\beta^2$ was applied to the visibility amplitudes of each source (see Equation~(\ref{eq:vis})). While there was no need to correct for Sgr A* due to its proximity to the maser source, $\beta^2 \sim$ 0.74 (i.e. $\beta \sim$ 0.86) was applied for NRAO 530 which is located at $\sim$15$^\circ$ higher elevation than the maser source (Table \ref{tab:image_result}). \\
This correction is closely related to the sky opacity due to the elevation difference between the maser source (i.e. Sgr B2 and OH 0.55-0.06) and the target (i.e. Sgr A* and NRAO 530) and can be written as,
\begin{eqnarray}
\label{eq:beta_anal}
\beta(\Delta e(t)) \propto e^{\frac{1}{2} [{\tau(t) - \tau(t')}]} = e^{\frac{1}{2} \tau_0 [sec(90^\circ - el(t))- sec(90^\circ - el(t'))]}.
\end{eqnarray}
where $\tau_0$ is opacity to the zenith, and $90^\circ-el(t)$ is the zenith angle, Z, at time $t$. 
The relationship of $\beta$ to the sec(Z) difference between the maser and target sources, sec(Z$_{\rm target}$) - sec(Z$_{\rm maser}$), and to the target source elevation is presented in Figure~\ref{fig:beta_elev}. 
On the left two panels of Figure~\ref{fig:beta_elev}, each colored point represents the $\beta$ values from the obtained gain solutions by the template spectrum method, and the trend of beta with different opacity to the zenith (i.e. $\tau_0$) is shown as the dashed and dash-dotted lines. The points out of the lines show the larger difference between the a-priori and template spectrum method (i.e. $\alpha$) by the relatively bad observing condition. 
Note that the shaded area in Figure~\ref{fig:beta_elev} (right panel) shows the elevation range of NRAO 530 during the observations, and the averaged $\beta$ for NRAO 530 within the $\tau_0$ range from 0.05 to 0.3 which are the minimum and maximum opacities in our measurements at all telescopes is 0.87 $\pm$ 0.09. This estimated $\beta$ value is well consistent with the $\beta$ obtained from Equation (\ref{eq:ampratio}). \\
%
%
Figure~\ref{fig:mas_dist} shows the known sky distribution of H$_2$O and SiO maser sources (\cite{valdettaro:2001, indermuehle:2013, yoon:2014}; Cho et al. in prep.). If we assume an arbitrary continuum target is correspond to the 2$^\circ$ x 2$^\circ$ pixel of all sky, $\sim$89 and $\sim$64 \% of targets can be found within 15$^\circ$ of separation angle from each maser source.
Thus, this implies that the template spectrum method can be clearly applied to more than two-third of all available targets and properly corrected through the gain ratio ($G^{AP}/G^{AC}$), or the intervening opacity measurements (e.g. every $\sim$2 hours) at each telescope. \\
%
%

\section{Conclusion and Summary \label{sec:conclusion}}
In this study, we investigated the amplitude calibrations for the EAVN through the quantitive differences between the a-priori and template spectrum methods using the KaVA and additional telescopes in Japan. As a result, both gain values and total flux densities of each source from each method were well consistent within 10\% when the maser lines are clearly detected insuring the high SNR of gain values from template spectrum method. However, the difference becomes larger at lower elevations, particularly below 10$^\circ$, so that it can be more strictly tested through the  observations which will be practically designed for the specific purpose in the future. \\
The template spectrum method has potential to correct the possible gain losses by pointing uncertainty during an observation and to extend the available telescopes for VLBI imaging applications, particularly in cases where telescopes cannot provide proper calibration files for amplitude (i.e. $T_{sys}^{*}$ and gain curve). Our observations showed image dynamic ranges were increased by 10-15\% by the additional telescopes (i.e. TAK and NRO). This is especially important for the Sgr A* studies where several maser sources are observable in a close region and hard to calibrate the amplitude due to its low elevations for the northern hemisphere arrays. Therefore, the template spectrum method is practically important not only for accurate amplitude calibration and comparison with a-priori method, but also extending KaVA to radio telescopes in East-Asia regions. 
We also demonstrated that the template spectrum method is applicable to distant sources (e.g. NRAO 530, $\sim$15$^\circ$ separation) by introducing a separation angle correction factor, $\beta$. Thus, based on the known maser source distribution on the sky (e.g. Figure~\ref{fig:mas_dist}), the template spectrum method can be more widely utilized, particularly for extragalactic sources, which are relatively difficult to find suitable maser sources compared to the galactic sources. \\
Our comparative study will also be useful to check the reliability of flux density and size measurements for a variety of sources, especially for long-term monitoring observations (particularly, KaVA large programs) where consistent data processing is strongly required. \\

\textbf{Funding} \\
This work was financially supported by the National Research Foundation of Korea (NRF) via Global PhD Fellowship Grant (I.C.; NRF-2015H1A2A1033752), and Korea Research Fellowship Program through the NRF funded by the Ministry of Science, ICT and Future Planning (G.-Y. Z. and T.J.; NRF-2015H1D3A1066561). K.A. is financially supported by the program of Postdoctoral Fellowships for Research Abroad at the Japan Society for the Promotion of Science and by a grant from the National Science Foundation (NSF; AST-1614868). The Black Hole Initiative at Harvard University is financially supported by a grant from the John Templeton Foundation.

\begin{ack}
We would like to thank the anonymous referee for the constructive comments and suggestions that further improved this paper. We also thank Mikyoung Kim for helpful suggestions, especially developing the sky distribution of maser sources. This work is based on observations made with the KaVA and JVN, which is operated by the the Korea Astronomy and Space Science Institute (KASI) and the National Astronomical Observatory of Japan (NAOJ). The Nobeyama 45-m radio telescope is operated by Nobeyama Radio Observatory (NRO), a branch of NAOJ. We are grateful to all staff members in KVN, VERA, Yamaguchi Radio Telescope, Takahagi Radio Telescope and NRO who helped to operate the array and to correlate the data. The KVN operations are supported by Korea Research Environment Open NETwork (KREONET) which is managed and operated by  Korea Institute of Science and Technology Information (KISTI). \\\\
\end{ack}

\setcounter{section}{0}
\renewcommand\thesection{}
\section{Appendix A \label{sec:appendix}}
%
\setcounter{table}{0}
\renewcommand{\thetable}{A\arabic{table}}
\begin{table}
\tbl{Typical uncertainties of measurements.}{%
\centering
\begin{tabular}{llllll}
\hline
Telescope &  ${\sigma_{A}/\eta_{A}}^{\rm a}$  &  ${\sigma_{T^*}/T_{sys}^{*}}^{\rm b}$  &  ${\eta_{\theta}}^{\rm c}$ & $\sigma_{i}^{AP}/G_{i}^{AP}$ & $\sigma_{i}^{AC}/G_{i}^{AC}$ \\
2013 Mar. & (22 GHz) &&&& \\
\hline
MIZ &  0.09 & 0.05-0.10 & 0.992 & 0.22 & 0.13 \\
IRK & 0.02 & 0.05-0.10 & 0.993 &0.09 & 0.13  \\
OGA& 0.03 & 0.05-0.10 & 0.987 & 0.11 & 0.13  \\
ISG & 0.06 & 0.05-0.10 & 0.991 & 0.16 & 0.13  \\
KYS & 0.04 & 0.05-0.10 & 0.996 & 0.12 & 0.13  \\ 
KUS & 0.03 & 0.05-0.10 & 0.996 & 0.12 & 0.12  \\
KTN & 0.03 & 0.05-0.10 & 0.994 & 0.13 & 0.13  \\
TAK & $<$0.33$^{\rm *}$  & 0.05-0.10 & 0.965 & $<$0.22$^{\rm *}$ & 0.13  \\
YAM & - & - & - & - & 0.13 \\
\hline
2013 Apr. & (43 GHz) &&&& \\
\hline
MIZ & 0.09 & 0.05-0.10 & 0.970 & 0.24 & 0.13  \\
IRK & 0.03 & 0.05-0.10 & 0.975 & 0.12 & 0.13   \\
OGA & 0.04 & 0.05-0.10 & 0.951 & 0.14 & 0.13  \\
ISG & 0.06 & 0.05-0.10 & 0.969 & 0.15 & 0.13  \\
KYS & 0.03 & 0.05-0.10 & 0.987 & 0.13 & 0.13  \\
KUS & 0.04 & 0.05-0.10 & 0.986 & 0.13 & 0.12  \\
KTN & 0.03 & 0.05-0.10 & 0.978 & 0.13 & 0.13  \\
NRO & 0.06 & 0.05-0.10 & 0.660 & 0.32 & 0.13 \\
\hline
2014 Nov. & (43 GHz) &&&& \\
\hline
MIZ & 0.05 & 0.05-0.10 & 0.970 & 0.17 & 0.13  \\
IRK & 0.02 & 0.05-0.10 & 0.975 & 0.09 & 0.13  \\
OGA & 0.06 & 0.05-0.10 & 0.951 & 0.18& 0.13  \\
ISG & 0.08 & 0.05-0.10 & 0.969 & 0.21 & 0.13  \\
KYS & 0.03 & 0.05-0.10 & 0.995 & 0.13 & 0.12  \\
KUS & 0.03 & 0.05-0.10 & 0.992 & 0.13 & 0.13  \\
KTN & 0.03 & 0.05-0.10 & 0.992 & 0.13 & 0.13  \\
\hline
\end{tabular}
}\label{tab:errors}
\begin{tabnote}
$^{\rm a}$ The aperture efficiency uncertainties, relative to the measured values in the KVN and VERA status reports in 2013 and 2014. \\
$^{\rm b}$ The typical system temperature measurement uncertainties of the chopper-wheel method (\cite{kutner:1981, jewell:2002}). \\
$^{\rm c}$ The gain efficiencies when the pointing offsets in a status report in 2013 and 2014 are given, relative to HPBW of the beam size. \\
$^{\rm *}$ Preliminary estimates (\cite{yonekura:2016}). \\
\end{tabnote}
\end{table}
%
The gain errors are mainly affected by the measurement uncertainties of aperture efficiency ($\sigma_{A}$), system temperature ($\sigma_{T}$) and the gain efficiency under a pointing offset ($\eta_{\theta}$) (see Table~\ref{tab:errors}). Considering each uncertainties, the gain solutions at $i$-th antenna at time $t$ from a-priori ($G_{i}^{AP}(t)$) and template spectrum ($G_{i}^{AC}(t)$) methods are represented with its errors ($\sigma_{i}^{AP}(t)$ and $\sigma_{i}^{AC}(t)$, respectively) as,
\setcounter{equation}{0}
\renewcommand{\theequation}{A\arabic{equation}}
\begin{equation}
\begin{array}{l}
\label{eq:gain_error}
\displaystyle G_{i}^{AP}(t) \pm \sigma_{i}^{AP}(t) \\
\displaystyle = \sqrt{\frac{1}{\eta_{\theta,i}} \frac{T_{sys,i}^{*}(t) \pm \sigma_{T,i}(t)}{DPFU_i  f_i(e(t)) \pm \sigma_{A,i}(e(t))}}, \\\\
\displaystyle G_{i}^{AC}(t) \pm \sigma_{i}^{AC}(t) \\
\displaystyle = \sqrt{ \frac{T_{sys,0}^{*}(t_0) \pm \sigma_{T,0}(t_0)}{DPFU_0  f_0(e(t_0)) \pm \sigma_{A,0}(e(t_0))} (\rho_{r,i}(t) \pm \sigma_{r,i}(t))}, \\
\end{array} 
\end{equation}

where $\sigma_{A,i}(e(t))$ and $\sigma_{T,i}(t)$ are the aperture efficiency and system temperature measurement uncertainties, respectively. $\rho_{r,i}(t)$ is the $\rho$-ratio (i.e. $\rho_0(t_0)/\rho_i(t)$) and its uncertainty is $\sigma_{r,i}(t)$ which is affected by the baseline fitting error and maser line selection in the total-power spectrum of a maser source. Note that $\rho_{r,i}(t)$ corrects the gain loss effects by pointing offset (i.e. $\eta_{\theta,i} = 1/\rm exp({4ln2(\theta_{off}/\theta_{hpbw})^2}$) where $\rm \theta_{off}$ and $\rm \theta_{hpbw}$ are pointing offset and HPBW of the beam, respectively; \cite{leej:2017}), and $\sigma_{r,i}(t)$ is typically less than $\sim$1 \% when the maser source and template spectrum are selected under the selection criteria (see Section~\ref{sec:templatemethod}). Therefore, the relative errors to each gain solution can be obtained by, 
\begin{equation}
\begin{array}{l}
\label{eq:relative_error}
\displaystyle \frac{\sigma_{i}^{AP}}{G_{i}^{AP}} \approx \sqrt{\frac{1}{\eta_{\theta,i}}(1 + \frac{2k}{A_i} \frac{\sigma_{A,i}}{\eta_{A,i}} + \frac{\sigma_{T,i}}{T_{sys,i}^{*}}) } - 1, \\\\
\displaystyle \frac{\sigma_{i}^{AC}}{G_{i}^{AC}} \approx \sqrt{1 + \frac{2k}{A_0} \frac{\sigma_{A,0}}{\eta_{A,0}} + \frac{\sigma_{T,0}}{T_{sys,0}^{*}} + \frac{\sigma_{r,i}}{\rho_{r,i}} } - 1.\\
\end{array} 
\end{equation}
%


\begin{thebibliography}{}
\expandafter\ifx\csname natexlab\endcsname\relax\def\natexlab#1{#1}\fi

\bibitem[{{Akiyama} {et~al.}(2013){Akiyama}, {Takahashi}, {Honma}, {Oyama}, \&
  {Kobayashi}}]{akiyama:2013}
{Akiyama}, K., {Takahashi}, R., {Honma}, M., {Oyama}, T., \& {Kobayashi}, H.
  2013, \pasj, 65, arXiv:1308.6657

\bibitem[{{Akiyama} {et~al.}(2014){Akiyama}, {Kino}, {Sohn}, {Lee}, {Trippe},
  \& {Honma}}]{akiyama:2014}
{Akiyama}, K., {Kino}, M., {Sohn}, B., {et~al.} 2014, in IAU Symposium, Vol.
  303, The Galactic Center: Feeding and Feedback in a Normal Galactic Nucleus,
  ed. L.~O. {Sjouwerman}, C.~C. {Lang}, \& J.~{Ott}, 288--292

\bibitem[{{Akiyama} {et~al.}(2015){Akiyama}, {Lu}, {Fish}, {Doeleman},
  {Broderick}, {Dexter}, {Hada}, {Kino}, {Nagai}, {Honma}, {Johnson}, {Algaba},
  {Asada}, {Brinkerink}, {Blundell}, {Bower}, {Cappallo}, {Crew}, {Dexter},
  {Dzib}, {Freund}, {Friberg}, {Gurwell}, {Ho}, {Inoue}, {Krichbaum},
  {Loinard}, {MacMahon}, {Marrone}, {Moran}, {Nakamura}, {Nagar}, {Ortiz-Leon},
  {Plambeck}, {Pradel}, {Primiani}, {Rogers}, {Roy}, {SooHoo}, {Tavares},
  {Tilanus}, {Titus}, {Wagner}, {Weintroub}, {Yamaguchi}, {Young}, {Zensus}, \&
  {Ziurys}}]{akiyama:2015}
{Akiyama}, K., {Lu}, R.-S., {Fish}, V.~L., {et~al.} 2015, \apj, 807, 150

\bibitem[{{Bower} {et~al.}(2015){Bower}, {Markoff}, {Dexter}, {Gurwell},
  {Moran}, {Brunthaler}, {Falcke}, {Fragile}, {Maitra}, {Marrone}, {Peck},
  {Rushton}, \& {Wright}}]{bower:2015}
{Bower}, G.~C., {Markoff}, S., {Dexter}, J., {et~al.} 2015, \apj, 802, 69

\bibitem[{{Doeleman} {et~al.}(2001){Doeleman}, {Shen}, {Rogers}, {Bower},
  {Wright}, {Zhao}, {Backer}, {Crowley}, {Freund}, {Ho}, {Lo}, \&
  {Woody}}]{doeleman:2001}
{Doeleman}, S.~S., {Shen}, Z.-Q., {Rogers}, A.~E.~E., {et~al.} 2001, \aj, 121,
  2610

\bibitem[{{Doeleman} {et~al.}(2008){Doeleman}, {Weintroub}, {Rogers},
  {Plambeck}, {Freund}, {Tilanus}, {Friberg}, {Ziurys}, {Moran}, {Corey},
  {Young}, {Smythe}, {Titus}, {Marrone}, {Cappallo}, {Bock}, {Bower},
  {Chamberlin}, {Davis}, {Krichbaum}, {Lamb}, {Maness}, {Niell}, {Roy},
  {Strittmatter}, {Werthimer}, {Whitney}, \& {Woody}}]{doeleman:2008}
{Doeleman}, S.~S., {Weintroub}, J., {Rogers}, A.~E.~E., {et~al.} 2008, \nat,
  455, 78

\bibitem[{{Doeleman} {et~al.}(2012){Doeleman}, {Fish}, {Schenck}, {Beaudoin},
  {Blundell}, {Bower}, {Broderick}, {Chamberlin}, {Freund}, {Friberg},
  {Gurwell}, {Ho}, {Honma}, {Inoue}, {Krichbaum}, {Lamb}, {Loeb}, {Lonsdale},
  {Marrone}, {Moran}, {Oyama}, {Plambeck}, {Primiani}, {Rogers}, {Smythe},
  {SooHoo}, {Strittmatter}, {Tilanus}, {Titus}, {Weintroub}, {Wright}, {Young},
  \& {Ziurys}}]{doeleman:2012}
{Doeleman}, S.~S., {Fish}, V.~L., {Schenck}, D.~E., {et~al.} 2012, Science,
  338, 355

\bibitem[{{Doi} {et~al.}(2006){Doi}, {Fujisawa}, {Harada}, {Nagayama},
  {Suematsu}, {Sugiyama}, {Habe}, {Honma}, {Kawaguchi}, {Kobayashi}, {Koyama},
  {Murata}, {Omodaka}, {Sorai}, {Sudou}, {Takaba}, {Takashima}, \&
  {Wakamatsu}}]{doi:2006}
{Doi}, A., {Fujisawa}, K., {Harada}, K., {et~al.} 2006, in Proceedings of the
  8th European VLBI Network Symposium, 71

\bibitem[{{Fish} {et~al.}(2011){Fish}, {Doeleman}, {Beaudoin}, {Blundell},
  {Bolin}, {Bower}, {Chamberlin}, {Freund}, {Friberg}, {Gurwell}, {Honma},
  {Inoue}, {Krichbaum}, {Lamb}, {Marrone}, {Moran}, {Oyama}, {Plambeck},
  {Primiani}, {Rogers}, {Smythe}, {SooHoo}, {Strittmatter}, {Tilanus}, {Titus},
  {Weintroub}, {Wright}, {Woody}, {Young}, \& {Ziurys}}]{fish:2011}
{Fish}, V.~L., {Doeleman}, S.~S., {Beaudoin}, C., {et~al.} 2011, \apjl, 727,
  L36

\bibitem[{{Greisen}(2003)}]{greisen:2003}
{Greisen}, E.~W. 2003, Information Handling in Astronomy - Historical Vistas,
  285, 109

\bibitem[{{Hada} {et~al.}(2016){Hada}, {Kino}, {Doi}, {Nagai}, {Honma},
  {Akiyama}, {Tazaki}, {Lico}, {Giroletti}, {Giovannini}, {Orienti}, \&
  {Hagiwara}}]{hada:2016}
{Hada}, K., {Kino}, M., {Doi}, A., {et~al.} 2016, \apj, 817, 131

\bibitem[{{Hagiwara} {et~al.}(2015){Hagiwara}, {An}, {Jung}, {Rho}, {Zhang},
  {Hao}, {Fujisawa}, {Yonekura}, {Baan}, {Kim}, \& {Kobayashi}}]{hagiwara:2015}
{Hagiwara}, Y., {An}, T., {Jung}, T., {et~al.} 2015, Publication of Korean
  Astronomical Society, 30, 641

\bibitem[{{H{\"o}gbom}(1974)}]{hogbom:1974}
{H{\"o}gbom}, J.~A. 1974, \aaps, 15, 417

\bibitem[{{Indermuehle} {et~al.}(2013){Indermuehle}, {Edwards}, {Brooks}, \&
  {Urquhart}}]{indermuehle:2013}
{Indermuehle}, B., {Edwards}, P., {Brooks}, K., \& {Urquhart}, J. 2013, in
  CSIRO Data Archive

\bibitem[{{Jewell}(2002)}]{jewell:2002}
{Jewell}, P.~R. 2002, in Astronomical Society of the Pacific Conference Series,
  Vol. 278, Single-Dish Radio Astronomy: Techniques and Applications, ed.
  S.~{Stanimirovic}, D.~{Altschuler}, P.~{Goldsmith}, \& C.~{Salter}, 313--328

\bibitem[{{Johnson} {et~al.}(2015){Johnson}, {Fish}, {Doeleman}, {Marrone},
  {Plambeck}, {Wardle}, {Akiyama}, {Asada}, {Beaudoin}, {Blackburn},
  {Blundell}, {Bower}, {Brinkerink}, {Broderick}, {Cappallo}, {Chael}, {Crew},
  {Dexter}, {Dexter}, {Freund}, {Friberg}, {Gold}, {Gurwell}, {Ho}, {Honma},
  {Inoue}, {Kosowsky}, {Krichbaum}, {Lamb}, {Loeb}, {Lu}, {MacMahon},
  {McKinney}, {Moran}, {Narayan}, {Primiani}, {Psaltis}, {Rogers}, {Rosenfeld},
  {SooHoo}, {Tilanus}, {Titus}, {Vertatschitsch}, {Weintroub}, {Wright},
  {Young}, {Zensus}, \& {Ziurys}}]{johnson:2015}
{Johnson}, M.~D., {Fish}, V.~L., {Doeleman}, S.~S., {et~al.} 2015, Science,
  350, 1242

\bibitem[{{Krichbaum} {et~al.}(1993{\natexlab{a}}){Krichbaum}, {Zensus},
  {Witzel}, {Mezger}, {Standke}, {Schalinski}, {Alberdi}, {Marcaide}, {Zylka},
  {Rogers}, {Booth}, {Ronnang}, {Colomer}, {Bartel}, \&
  {Shapiro}}]{krichbaum:1993a}
{Krichbaum}, T.~P., {Zensus}, J.~A., {Witzel}, A., {et~al.} 1993{\natexlab{a}},
  \aap, 274, L37

\bibitem[{{Krichbaum} {et~al.}(1993{\natexlab{b}}){Krichbaum}, {Witzel},
  {Graham}, {Standke}, {Schwartz}, {Lochner}, {Schalinski}, {Greve}, {Steppe},
  {Brunswig}, {Butin}, {Hein}, {Navarro}, {Penalver}, {Grewing}, {Booth},
  {Colomer}, \& {Ronnang}}]{krichbaum:1993b}
{Krichbaum}, T.~P., {Witzel}, A., {Graham}, D.~A., {et~al.} 1993{\natexlab{b}},
  \aap, 275, 375

\bibitem[{{Kutner} \& {Ulich}(1981)}]{kutner:1981}
{Kutner}, M.~L., \& {Ulich}, B.~L. 1981, \apj, 250, 341

\bibitem[{{Lee} {et~al.}(2016){Lee}, {Lee}, {Kang}, {Byun}, \&
  {Kim}}]{leej:2016}
{Lee}, J.~W., {Lee}, S.-S., {Kang}, S., {Byun}, D.-Y., \& {Kim}, S.~S. 2016,
  \aap, 592, L10

\bibitem[{{Lee} {et~al.}(2017){Lee}, {Sohn}, {Byun}, {Lee}, \&
  {Kim}}]{leej:2017}
{Lee}, J.~W., {Sohn}, B.~W., {Byun}, D.-Y., {Lee}, J.~A., \& {Kim}, S.~S. 2017,
  ArXiv e-prints, arXiv:1703.05894

\bibitem[{{Lee} {et~al.}(2015{\natexlab{a}}){Lee}, {Oh}, {Roh}, {Oh}, {Kim},
  {Yeom}, {Kim}, {Jung}, {Byun}, {Jung}, {Kawaguchi}, {Shibata}, \&
  {Wajima}}]{lee:2015a}
{Lee}, S.-S., {Oh}, C.~S., {Roh}, D.-G., {et~al.} 2015{\natexlab{a}}, Journal
  of Korean Astronomical Society, 48, 125

\bibitem[{{Lee} {et~al.}(2015{\natexlab{b}}){Lee}, {Byun}, {Oh}, {Kim}, {Kim},
  {Jung}, {Oh}, {Roh}, {Jung}, \& {Yeom}}]{lee:2015b}
{Lee}, S.-S., {Byun}, D.-Y., {Oh}, C.~S., {et~al.} 2015{\natexlab{b}}, Journal
  of Korean Astronomical Society, 48, 229

\bibitem[{{Lu} {et~al.}(2011){Lu}, {Krichbaum}, {Eckart}, {K{\"o}nig},
  {Kunneriath}, {Witzel}, {Witzel}, \& {Zensus}}]{lu:2011}
{Lu}, R.-S., {Krichbaum}, T.~P., {Eckart}, A., {et~al.} 2011, \aap, 525, A76

\bibitem[{{Mart{\'{\i}}-Vidal} {et~al.}(2012){Mart{\'{\i}}-Vidal}, {Krichbaum},
  {Marscher}, {Alef}, {Bertarini}, {Bach}, {Schinzel}, {Rottmann}, {Anderson},
  {Zensus}, {Bremer}, {Sanchez}, {Lindqvist}, \& {Mujunen}}]{marti-vidal:2012}
{Mart{\'{\i}}-Vidal}, I., {Krichbaum}, T.~P., {Marscher}, A., {et~al.} 2012,
  \aap, 542, A107
  
\bibitem[{{McGrath} {et~al.}(2004){McGrath}, {Goss}, \& {De
  Pree}}]{mcgrath:2004}
{McGrath}, E.~J., {Goss}, W.~M., \& {De Pree}, C.~G. 2004, \apjs, 155, 577

\bibitem[{{Miyazaki} \& {Kobayashi}(2009)}]{miyazaki:2009}
{Miyazaki}, A., \& {Kobayashi}, H. 2009, in New Science Enabled by
  Microarcsecond Astrometry, held 21-23 July 2009 in Socorro, NM. 
  
\bibitem[{{Moran} \& {Dhawan}(1995)}]{moran:1995}
{Moran}, J.~M., \& {Dhawan}, V. 1995, in Astronomical Society of the Pacific
  Conference Series, Vol.~82, Very Long Baseline Interferometry and the VLBA,
  ed. J.~A. {Zensus}, P.~J. {Diamond}, \& P.~J. {Napier}, 161

\bibitem[{{Niinuma} {et~al.}(2014){Niinuma}, {Lee}, {Kino}, {Sohn}, {Akiyama},
  {Zhao}, {Sawada-Satoh}, {Trippe}, {Hada}, {Jung}, {Hagiwara}, {Dodson},
  {Koyama}, {Honma}, {Nagai}, {Chung}, {Doi}, {Fujisawa}, {Han}, {Kim}, {Lee},
  {Lee}, {Miyazaki}, {Oyama}, {Sorai}, {Wajima}, {Bae}, {Byun}, {Cho}, {Choi},
  {Chung}, {Chung}, {Han}, {Hirota}, {Hwang}, {Je}, {Jike}, {Jung}, {Jung},
  {Kang}, {Kang}, {Kang}, {Kan-ya}, {Kanaguchi}, {Kawaguchi}, {Kim}, {Kim},
  {Kim}, {Kim}, {Kim}, {Kim}, {Kim}, {Kobayashi}, {Kono}, {Kurayama}, {Lee},
  {Lee}, {Lee}, {Minh}, {Matsumoto}, {Nakagawa}, {Oh}, {Oh}, {Park}, {Roh},
  {Sasao}, {Shibata}, {Song}, {Tamura}, {Wi}, {Yeom}, \& {Yun}}]{niinuma:2014}
{Niinuma}, K., {Lee}, S.-S., {Kino}, M., {et~al.} 2014, \pasj, 66, 103

\bibitem[{{Oyama} {et~al.}(2008){Oyama}, {Miyoshi}, {Deguchi}, {Imai}, \&
  {Shen}}]{oyama:2008}
{Oyama}, T., {Miyoshi}, M., {Deguchi}, S., {Imai}, H., \& {Shen}, Z.-Q. 2008,
  \pasj, 60, 11

\bibitem[{{Rauch} {et~al.}(2016){Rauch}, {Ros}, {Krichbaum}, {Eckart},
  {Zensus}, {Shahzamanian}, \& {Mu{\v z}i{\'c}}}]{rauch:2016}
{Rauch}, C., {Ros}, E., {Krichbaum}, T.~P., {et~al.} 2016, \aap, 587, A37

\bibitem[{{Reid} \& {Brunthaler}(2004)}]{reid:2004}
{Reid}, M.~J., \& {Brunthaler}, A. 2004, \apj, 616, 872

\bibitem[{{Shen} {et~al.}(2005){Shen}, {Lo}, {Liang}, {Ho}, \&
  {Zhao}}]{shen:2005}
{Shen}, Z.-Q., {Lo}, K.~Y., {Liang}, M.-C., {Ho}, P.~T.~P., \& {Zhao}, J.-H.
  2005, \nat, 438, 62

\bibitem[{{Shepherd}(1997)}]{shepherd:1997}
{Shepherd}, M.~C. 1997, in Astronomical Society of the Pacific Conference
  Series, Vol. 125, Astronomical Data Analysis Software and Systems VI, ed.
  G.~{Hunt} \& H.~{Payne}, 77

\bibitem[{{Tsuboi} {et~al.}(2015){Tsuboi}, {Asaki}, {Kameya}, {Yonekura},
  {Miyamoto}, {Kaneko}, {Seta}, {Nakai}, {Takaba}, {Wakamatsu}, {Miyoshi},
  {Fukuzaki}, {Uehara}, \& {Sekido}}]{tsuboi:2015}
{Tsuboi}, M., {Asaki}, Y., {Kameya}, O., {et~al.} 2015, \apjl, 798, L6

\bibitem[{{Valdettaro} {et~al.}(2001){Valdettaro}, {Palla}, {Brand},
  {Cesaroni}, {Comoretto}, {Di Franco}, {Felli}, {Natale}, {Palagi}, {Panella},
  \& {Tofani}}]{valdettaro:2001}
{Valdettaro}, R., {Palla}, F., {Brand}, J., {et~al.} 2001, \aap, 368, 845

\bibitem[{{Wajima} {et~al.}(2016){Wajima}, {Hagiwara}, {An}, {Baan},
  {Fujisawa}, {Hao}, {Jiang}, {Jung}, {Kawaguchi}, {Kim}, {Kobayashi}, {Oh},
  {Roh}, {Wang}, {Xia}, \& {Zhang}}]{wajima:2016}
{Wajima}, K., {Hagiwara}, Y., {An}, T., {et~al.} 2016, in Astronomical Society
  of the Pacific Conference Series, Vol. 502, Frontiers in Radio Astronomy and
  FAST Early Sciences Symposium 2015, ed. L.~{Qain} \& D.~{Li}, 81

\bibitem[{{Yonekura} {et~al.}(2016){Yonekura}, {Saito}, {Sugiyama}, {Soon},
  {Momose}, {Yokosawa}, {Ogawa}, {Kimura}, {Abe}, {Nishimura}, {Hasegawa},
  {Fujisawa}, {Ohyama}, {Kono}, {Miyamoto}, {Sawada-Satoh}, {Kobayashi},
  {Kawaguchi}, {Honma}, {Shibata}, {Sato}, {Ueno}, {Jike}, {Tamura}, {Hirota},
  {Miyazaki}, {Niinuma}, {Sorai}, {Takaba}, {Hachisuka}, {Kondo}, {Sekido},
  {Murata}, {Nakai}, \& {Omodaka}}]{yonekura:2016}
{Yonekura}, Y., {Saito}, Y., {Sugiyama}, K., {et~al.} 2016, \pasj, 68, 74

\bibitem[{{Yoon} {et~al.}(2014){Yoon}, {Cho}, {Kim}, {Yun}, \&
  {Park}}]{yoon:2014}
{Yoon}, D.-H., {Cho}, S.-H., {Kim}, J., {Yun}, Y.~j., \& {Park}, Y.-S. 2014,
  \apjs, 211, 15

\bibitem[{{Zhao} {et~al.}(2017){Zhao}, {Kino}, {Cho}, {Akiyama}, {Sohn},
  {Jung}, {Algaba}, {Hada}, {Hagiwara}, {Hodgson}, {Honma}, {Kawaguchi},
  {Koyama}, {Lee}, {Lee}, {Niinuma}, {Oh}, {Park}, {Ro}, {Sawada-Satoh},
  {Tazaki}, {Trippe}, {Wajima}, \& {Yoo}}]{zhao:2017}
{Zhao}, G.-Y., {Kino}, M., {Cho}, I.-J., {et~al.} 2017, in IAU Symposium, Vol.
  322, IAU Symposium, ed. R.~M. {Crocker}, S.~N. {Longmore}, \& G.~V.
  {Bicknell}, 56--63

\bibitem[{{Zheng}(2015)}]{zheng:2015}
{Zheng}, W. 2015, IAU General Assembly, 22, 2255896

\end{thebibliography}
\end{document}